# Geo: A Query Rewrite Framework for Graph Pattern Mining

NAZANIN YOUSEFIAN, Simon Fraser University, Canada
KASRA JAMSHIDI, Simon Fraser University, Canada
KEVAL VORA, Simon Fraser University, Canada
ANDERS MILTNER, Simon Fraser University, Canada

Graph pattern mining is important for analyzing graph data. Graph mining systems typically require answering pattern matching queries, which involve solving the NP-complete subgraph isomorphism problem. To address this, domain experts often develop custom pattern matching query optimization strategies based on exploiting substructural similarities across different patterns. While these optimizers can be effective, their development is challenging due to the complex structural properties of the patterns (e.g., subsymmetries), which are difficult to address. This complexity limits the exploration of interactions between different optimization strategies and restricts experts from continuously improving the optimizers—such as by incorporating additional custom or general pattern-based equivalences over time.

In this paper, we present a programmable pattern matching query optimizer called Geo, which automatically manages the interactions between various equivalences, ensures the optimizations maintain correctness of results, and simplifies the management of substructure equivalences. Geo exposes a simple but flexible language for expressing pattern equivalences as rewrite rules. By maintaining canonical representations of generated patterns during equality saturation, Geo avoids issues arising from syntactic differences in isomorphic patterns. Additionally, we develop *embedded reconstructablility* (EmRec) that tracks provenance across equivalences to ensure various reconstructability needs of desired outputs. Our evaluation demonstrates that Geo can discover novel query equivalences through complex composition of various rewrite rules, enabling our optimized queries to achieve a cost reduction of up to 99% compared to the queries in prior work. We further test Geo's effectiveness at speeding up practical graph mining problems by using it in two representative case studies – approximate pattern matching and quasi-clique mining, and find it is highly effective at optimizing these tasks, enabling cost reductions of up to 71%.

## 1 Introduction

Graph mining is the process of querying substructures of interest within large data graphs. Graph mining is important across several domains like social network analysis, security and bioinformatics [Milo et al. 2002; Rotabi et al. 2017; Sarkar et al. 2019; Wang et al. 2019]. Systems for graph

Authors' Contact Information: Nazanin Yousefian, Simon Fraser University, Burnaby, BC, Canada, nazanin_yousefian@sfu.ca; Kasra Jamshidi, Simon Fraser University, Burnaby, BC, Canada, kjamshid@cs.sfu.ca; Keval Vora, Simon Fraser University, Burnaby, BC, Canada, keval@cs.sfu.ca; Anders Miltner, Simon Fraser University, Burnaby, BC, Canada, miltner@cs.sfu.ca.

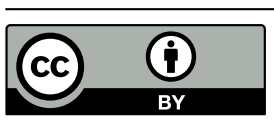

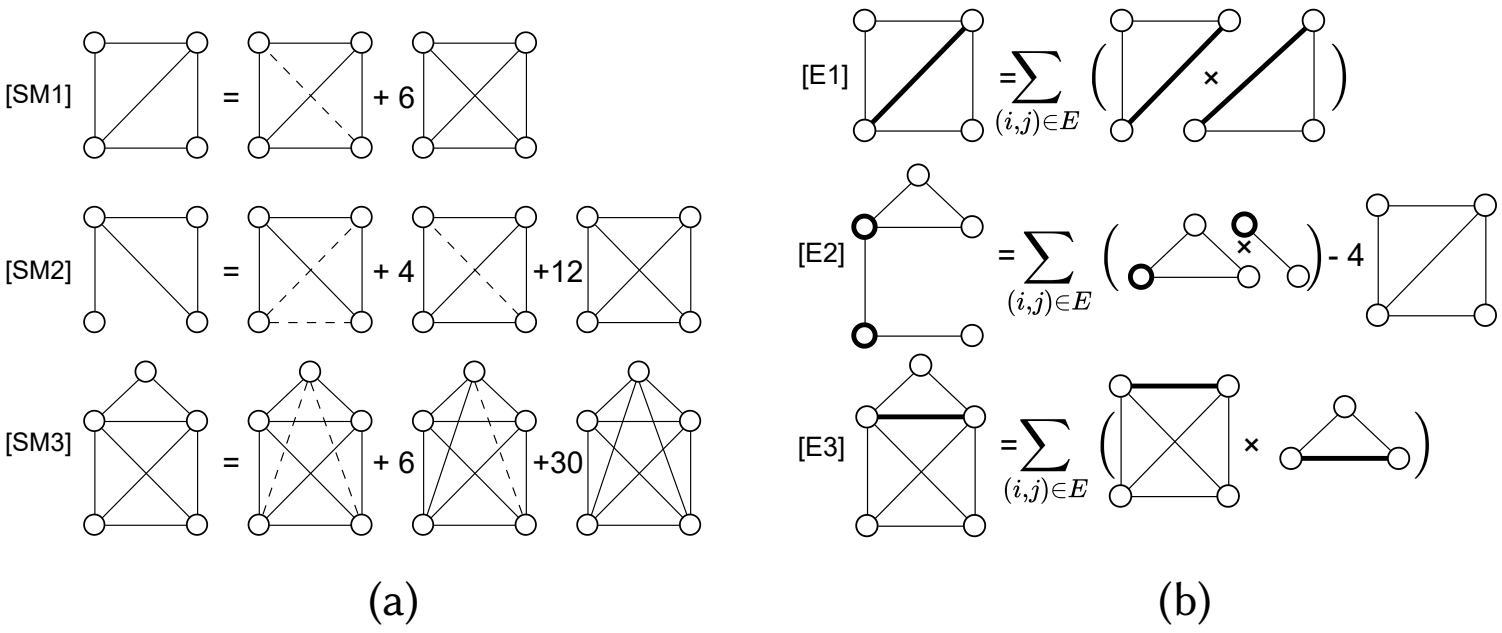

Fig. 1. Example rewrite rules in Subgraph Morphing (a) and ESCAPE (b). Dashed lines represent anti-edges.

mining are primarily based on solving the subgraph isomorphism problem. In essence, by characterizing these substructures of interest as subgraphs, graph mining systems solve the subgraph isomorphism problem. Subgraph isomorphism is an NP-complete problem, developing efficient subgraph matching algorithms is an active research area [Ammar et al. 2018; Bhattarai et al. 2019; Bi et al. 2016; Han et al. 2019, 2013; Kim et al. 2016; Lai et al. 2015, 2016; Mhedhbi and Salihoglu 2019; Qiao et al. 2017; Ren et al. 2019; Shao et al. 2014; Sun and Luo 2020; Yang et al. 2021a] as is building high-performance graph mining systems that incorporate efficient pattern mining techniques and optimizations [Che et al. 2024; Chen and Qian 2022, 2023; Chen and Arvind 2022; Chen et al. 2021; Gui et al. 2023; Jamshidi et al. 2020; Jamshidi and Vora 2021].

Recent works [Jamshidi et al. 2023] have shown that the performance of graph mining is not only sensitive to the pattern matching techniques being employed, but also to the input query patterns and the data graph – for example some data graphs might allow matching the triangle pattern faster, and some data graphs might allow matching the square pattern faster. Hence, different graph mining queries may run faster or slower depending on which patterns are queried, even if the results are equivalent.

Based on this insight, works like Subgraph Morphing [Jamshidi et al. 2023] leverage the substructural similarities across different patterns to optimize graph mining systems. In short, instead of simply mining the query patterns, the key idea here is to first transform those query patterns into certain alternative patterns (based on the substructural equivalences) that are relatively inexpensive to mine. This line of optimization is effective as it is orthogonal to the subgraph matching algorithm, and can be plugged in as an optimizer for graph mining systems that are based on any subgraph matching algorithm.

These kinds of optimizations often result into pattern-level equivalences developed by domain experts. For example, in Subgraph Morphing, equivalences are identified between subgraphs with same number of vertices. Hence, as shown in Figure 1(a)[SM2], the count for a tailed triangle can be derived from the counts of the three patterns on the right-hand side: the same pattern with anti-edges [1], a diamond with an anti-edge, and a 4-clique pattern.

On the other hand, ESCAPE [Pinar et al. 2017] decomposes query patterns into smaller patterns and this decomposition technique can also be viewed as pattern-level equivalences. For example, as shown in Figure 1(b)[E3], the count for the pattern on the left can be computed by identifying edges that are part of both a 4-clique and a triangle.

However, building a graph mining optimizer with such complex pattern-level equivalences is not easy as it requires addressing multiple problems that all require domain expertise. There are nuances in how pattern-level equivalences can be correctly applied, how different kinds of equivalences interact with each other, and how the potential explosion of search space of compositional equivalences can be managed. With the lack of a general framework to simplify expression of pattern equivalences, domain experts need to manually code the equivalences and their custom interactions, which is both tedious and error-prone. This limits domain experts from continuously improving the optimizers – such as by incorporating additional equivalences over time.

[1]An Anti-Edge enforces strict disconnection between two vertices in the subgraph [Jamshidi et al. 2020].

We aim to help subgraph isomorphism experts to focus on identifying interesting equivalences between pattern matching queries without worrying about such low-level details. In this work, we build a programmable graph mining optimizer called GEO that manages the interactions between equivalences using E-graphs [Nelson and Oppen 1980]. E-graphs are a popular data structure that, given a set of equivalences written as rewrite rules, and an input expression, can precisely represent the space of equivalent programs. Hence, with the set of pattern equivalence rewrite rules specified in our language, for a candidate pattern matching query, we can traverse the equivalences to find a candidate query that is more efficient than the input one. Recent advances in e-graph technology [Willsey et al. 2021] enable efficient composition of rewrite rules. However, using e-graphs to traverse such equivalences requires addressing two challenges unique to the pattern mining domain. First, patterns are embedded within pattern mining queries, and patterns have a rich set of equivalences not well-suited to a simple equational theory, creating a large search space; we must find a way to efficiently traverse such equivalences. Second, pattern mining queries are often run in batches; we must find a way to jointly optimize batches of queries to enable sharing sub-results.

*Pattern Equivalences.* E-graphs are syntax-based and do not natively understand pattern structures. In the context of e-graphs, two patterns that are isomorphic but written in different manners would not be considered equivalent. Naïvely addressing this by adding in rewrite rules according to graph equivalences would result in exponential explosion of search space primarily filled with symmetric patterns. To avoid this, we utilize existing work on *graph canonicalization* to find a canonical representation for each pattern maintained in the e-graph and across all rewrite rules. Then, as long as the provided rewrite rules *respect* the underlying canonicalization, traversing only canonicalized terms is sufficient for ensuring correctness of the optimization algorithm. We formalize this notion of rewrites respecting equivalence, and identify a simple syntactic condition that is sufficient to ensure that a rewrite respects equivalences.

*Joint Optimization of Pattern Batches.* Often graph analysts are interested in learning multiple patterns, and it is advantageous for a system to handle all such patterns in one shot, enabling joint optimizations [Jamshidi et al. 2023]. However, various graph mining problems have different requirements around how they translate to pattern matching queries and how the results get reconstructed from the pattern matching results. In such cases, depending on the *reconstructibility* requirements, the graph mining system must either be able to reconstruct complete rewrites for each individual query pattern in the query batch, or reconstruct rewrites for the entire query batch as a whole. As equivalences across multiple patterns are combined to explore more optimal rewrites, such combined exploration requires domain-specific reasoning principles specific to the reconstructibility needs of the graph mining problem.

However, expressing reconstructability in batched queries in a way that is amenable to e-graph optimization is tricky. We address this problem through *embedded reconstructablility* (EMREC). Through EMREC, GEO keeps track of the uses of each individual pattern via lightweight usage annotations. Inspired by data provenance [Buneman et al. 2001], our *reconstruction paths* keep track of how and why certain patterns are used, ensuring that patterns are not spuriously dropped when their usage in one query in the match may seemingly cancel out with another. By specifying how reconstruction paths are retained as equivalence rules interact with each other, the reconstruction paths capture what portions of newly generated queries are needed for reconstructing the original desired outputs. These reconstructability paths are more than an optimization – they enable data analysts to design custom notions of reconstructability.

We combined these developments in the Geo framework. Geo can be instantiated with rewrite rules capturing different pattern-level equivalences. We instantiated Geo with two classes of rewrite rules: one from Subgraph Morphing [Jamshidi et al. 2023], which is a generic algebra over subgraph structures and can be invoked on any arbitrary pattern. Second is a set of 15 custom rules from ESCAPE [Pinar et al. 2017] that are applied on specific size-4 and size-5 patterns. Hence, Geo is unique in the domain of graph mining optimizers, as it can be used to optimize arbitrary query patterns while also automatically leveraging interactions between different kinds of pattern-based equivalences resulting in novel equivalences.

We thoroughly evaluated Geo by studying its effectiveness in optimizing a wide range of pattern matching queries, consisting of commonly used patterns in previous graph mining systems research. We found that Geo was able to discover novel optimizations, enabling it to find queries that can run faster when compared to existing work. We then ablated the reconstructibility requirements captured via different EmRec reconstruction paths, and observed that Geo found optimal solutions that are custom to the specified reconstructibility requirement. Finally, we ablate canonicalization, and found that it enables faster identification of lower-cost queries than a non-canonicalized strawman.

Furthermore, we conducted two case studies: Approximate Pattern Mining and Quasi-Clique Mining. These mining tasks are fundamental across various domains like bioinformatics, social analysis, and fraud/crime detection [Brunato et al. 2007; Iyer et al. 2018]. These mining tasks get translated to pattern matching queries with different reconstructibility requirements. We observed that Geo generated equivalent optimized queries that are up to 64% to 71% more efficient on Approximate Pattern Mining and Quasi-Clique Mining, respectively.

After providing an overview of our system and some extended examples (§2), we present the following contributions:

- We provide a language for pattern matching queries that is amenable to rewrite-based optimizations (§3). Despite being relatively simple, this language supports flexible notions of reconstructability through its reconstruction-centered semantics, while enabling safe batching optimizations through EmRec.
- We describe how to encode rewrite-based query optimizations that work with our optimization algorithm (§4). Due to our canonicalization-centered approach, our algorithm is quite efficient despite the rich set of equivalences on queries.
- We formalize the pattern query optimization problem and provide an algorithm that performs this optimization (§5). Because we require our rewrites respect the canonicalizations, we are able to use state-of-the-art e-graph libraries like egg to efficiently optimize our queries.
- We provide Geo, an implementation of this language and optimization framework (§6).
- We evaluate the effectiveness of Geo (§7). We find that Geo outperforms prior work, optimizing queries by up to 99%. We ablate our usage of flexible reconstructions, and find that, in addition to enabling more queries, they enable optimizations improving queries by up to 92%. We ablate our optimization of operating over canonicalized forms, and find that the non-canonical form performs 13.5× worse than the full tool.
- We demonstrate the effectiveness of Geo at addressing two well-established graph mining tasks: approximate pattern mining and quasi-clique mining (§8). We find that Geo is effective at optimizing the pattern matching queries needed to solve these tasks, enabling an improvement of up to 71%.

## 2 Overview

In this section we provide an overview of the full GEO system. We first describe some example graph mining problems, their analogous pattern matching queries, and the goals of our full system. Then, we describe our solution, GEO, which includes a core language for modeling graph mining problems as pattern matching queries, and our rewrite-based query optimizer.

### 2.1 Graph Mining Problems and Pattern Matching Queries

Graph Mining Problems aim to analyze subgraphs of interests in large graphs. They do so by solving the subgraph isomorphism problem, which finds subgraphs in the given data graph that match small input pattern graphs.

There are various graph mining problems, targeting different objectives. For example:

- SUBGRAPH MINING finds subgraphs that match a specific pattern or a set of patterns, which is useful in applications like anomaly detection and fraud detection where anomalous/fraudulent patterns are well-defined.
- MOTIF COUNTING computes the distribution of recurring subgraph patterns of a particular size, providing valuable insights for exploratory data analysis by revealing the underlying structural characteristics and patterns within graphs.
- APPROXIMATE PATTERN MINING finds subgraphs that are structurally similar to (even though not isomorphic to) a specific pattern. This is useful in finding relevant subgraphs in noisy graphs and in scenarios where exact patterns are deemed to be too restrictive.
- QUASI-CLIQUE MINING finds dense subgraphs, which is useful in detecting small tightly-knit communities and anomalies with strong relationships.

Graph mining problems are often expressed as Pattern Matching Queries, allowing graph mining developers to tap into the advances in efficient subgraph isomorphism techniques. For instance, recent graph mining systems [Che et al. 2024; Chen et al. 2021; Jamshidi et al. 2020] employ pattern matching engines that efficiently explore all subgraphs that are isomorphic to a given pattern or a set of patterns. By doing so, the graph mining problem is solved by simply translating it into pattern matching queries that are passed to the pattern matching engine. Hence, for example, the motif counting problem boils down to a pattern matching query where the set of query patterns is all patterns of a specific size. On the other hand, the pattern matching query for the approximate pattern mining problem includes query patterns that are structurally similar to the given specific pattern. As expected, the subgraph mining problem translates directly to a pattern matching query.

#### 2.1.1 *Requirements of Graph Mining and Pattern Matching.*

We model the requirements of graph mining problems and pattern matching queries, which will form the basis for our solution. There are three key requirements.

*Pattern Grouping* (SINGLE *versus* BATCHED). Graph mining problems could translate to pattern matching queries that either match a single pattern (we call this SINGLE pattern query) or any of the patterns from a given set (we call this BATCHED pattern query). While single pattern queries can be viewed as special cases of batched pattern queries, we distinguish the two since batched pattern query often allows optimizing the pattern matching process by exploiting structural similarities to reduce redundancies in exploration (explained in Section 2.1.2). However, this requires some finesse; our language must be sufficiently expressive to enable sharing between subcomputations.

*Result Reconstructability* (INDIVIDUAL *versus* COLLECTIVE). For batched pattern queries, the requirements for the reconstructability of results to patterns can be categorized into two types. First is the INDIVIDUAL reconstructability, where matching results are aggregated based on each individual pattern in the query pattern batch. For example, the motif counting problem demands individual reconstructability in order to gather subgraph counts for every individual pattern separately, which

$$p = p^V + \sum_{q \supset p} q^V \times \phi(p, q)$$

Fig. 2. Generalized rewrite rule of Subgraph Morphing. For any pattern $x$, $x^V$ denotes the pattern with anti-edges added to each disconnected edge. The term $q \supset p$ means that $q$ is a super-pattern of $p$ with the same number of vertices. The function $\phi(p, q)$ represents the number of isomorphisms of $p$ within $q$.

together represents their overall distribution. On the other hand, Collective reconstructability requires the subgraph results to be aggregated together regardless of the query pattern they match. For example, for approximate pattern mining, the expected result set includes a single count value that indicates the number of subgraphs that match any of the patterns in the pattern query batch. While the collective reconstructability requirement can be easily satisfied by breaking the problem down into multiple individual reconstructability problems and combining their results into one (e.g., summing together subgraph counts for each individual pattern), we distinguish the two because collective reconstructability again allows optimizing the pattern matching process by reducing redundancies in mining similar subgraphs (explained in Section 2.1.2).

*Performance.* The subgraph isomorphism problem is NP-complete, and hence, the performance of executing the pattern matching queries is an important factor for graph mining problems. There are two broad ways to improve performance of graph mining problems. First, by speeding up the subgraph isomorphism algorithm in the pattern matching phase via efficient matching strategies customized for the given patterns in the query batch. And second, by rewriting pattern matching queries into other equivalent pattern matching queries that are faster to compute. The two ways are orthogonal since rewriting pattern matching queries is often informed by cost models that capture the performance effects of the matching strategies for any given query pattern. In this work, we focus on rewriting pattern matching queries to generate high-performance graph matching queries.

*2.1.2 Examples of Pattern Matching Rewrite Rules.* We discuss two specific rewrite rules here, highlighting key differences in the types of correlations across patterns they leverage, as well as their mathematical expressions.

Subgraph Morphing *Rewrite Rules.* The rewrite rules in Subgraph Morphing [Jamshidi et al. 2023] leverage super-pattern/sub-pattern relationships. These rules generate super-patterns from the original query patterns, ensuring that each super-pattern contains the same number of vertices as the query. The generated super-patterns are pairwise disjoint, and results are computed accounting for the isomorphisms of the query pattern within super-patterns, yielding integer coefficients. Figure 2 shows the general Subgraph Morphing equation which can be used for arbitrary patterns.

Figure 1(a) illustrates examples of rewrite rules derived from the general equation. For instance, the counts of a size-4 diamond pattern in equation [SM1] can be determined by counting the two super-patterns shown on the right and multiplying the count of cliques by 6.

ESCAPE *Rewrite Rules.* The rewrite rules in ESCAPE [Pinar et al. 2017] decompose query patterns into smaller, simpler sub-patterns. This approach accelerates pattern mining by counting occurrences of these sub-patterns in the data graph individually and then aggregating the results, while eliminating overlaps or common elements.

The ESCAPE rewrite rules are tailored specifically for size-4 and size-5 query patterns. Figure 1(b) illustrates examples of these rewrite rules. For instance, in Figure 1(b) [E1], instead of directly exploring the size-4 diamond pattern shown on the left, the counts can be obtained by identifying all edges that are part of two triangles, which is faster to compute.

Several derived formulae are obtained from the general rules in ESCAPE for all isomorphic variations of size-4 and size-5 patterns, with the exception of cliques (graphs where all vertices are fully connected) and cycles (closed-loop structures).

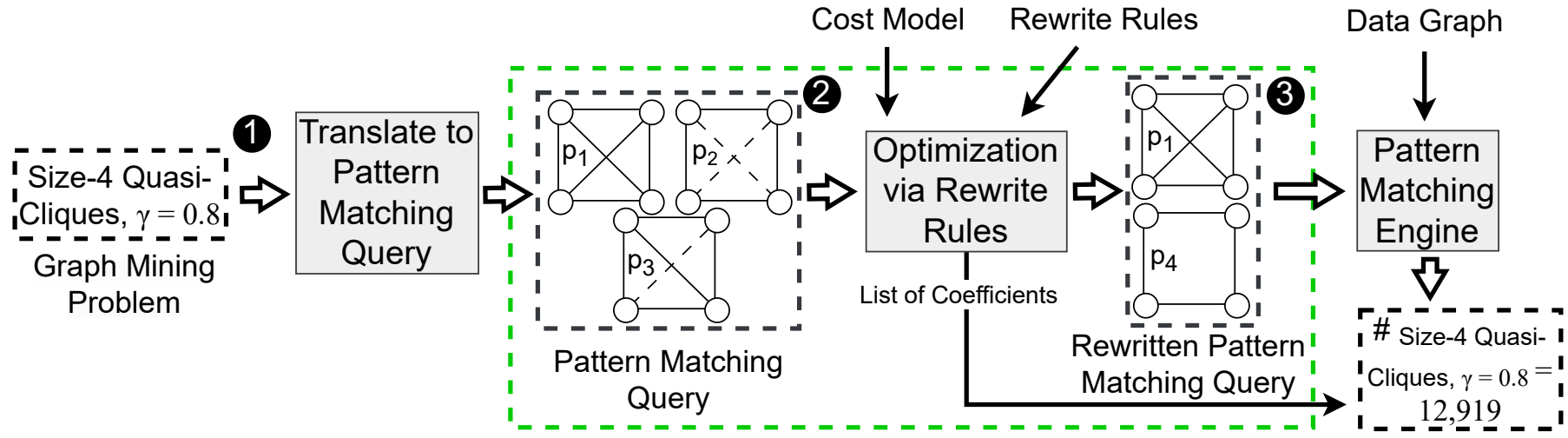

Fig. 3. Quasi-clique mining example, $\gamma = 0.8$

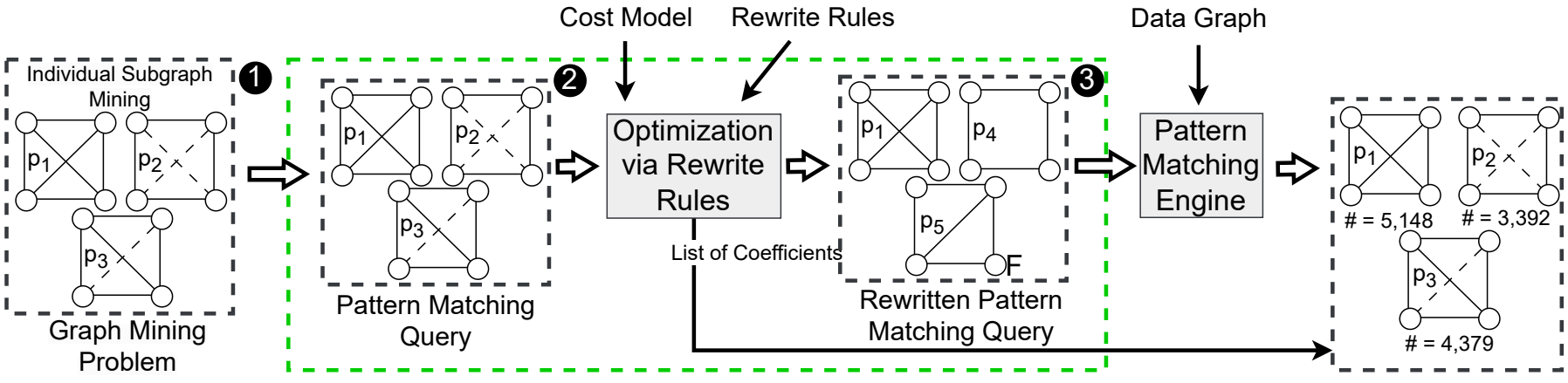

Fig. 4. Subgraph mining example.

*2.1.3 End-to-End Examples of Graph Mining & Pattern Matching Queries.* We present two examples of graph mining problems, focusing on their translation to pattern matching queries, which are then optimized via rewrite rules to generate equivalent pattern matching queries for the downstream pattern matching engine. To showcase the key differences in various requirements, we show: (a) an example of a batched graph mining problem with collective reconstructability (Quasi-Clique Mining in Figure 3); and, (b) another example of a batched graph mining problem with individual reconstructability (Subgraph Mining Problem in Figure 4).

*Quasi-Clique Mining (Batched, Collective).* Figure 3 shows quasi-clique mining which aims to find the number of all dense subgraphs with 4 vertices where density is captured by $\gamma$, i.e., each vertex of the subgraph is connected to at least $3 \times \gamma$ vertices (box #1). This results into batched pattern matching query with collective reconstructability requirement containing the three patterns shown inside box #2. Using a cost model, this batch of three patterns from box #2 is then rewritten into an equivalent pattern matching query containing low-cost patterns (i.e., patterns that are faster to mine) shown in box #3. The pattern matching engine efficiently explores the subgraphs that match either of the patterns from box #3, and returns the result, which in this case is a single count indicating the number of subgraphs qualified as quasi-cliques.

*Subgraph Mining (Batched, Individual).* Figure 4 shows the subgraph mining example which aims to individually compute the number of occurrences of each of the patterns in the input pattern batch (box #1). The translation to pattern matching query is an identity function in this case, and hence box #2 shows the same three input patterns in the batch. The pattern matching query gets rewritten into an equivalent query with a batch of three patterns (box #3) that are expected to be faster to compute based on the cost model. Even though box #2 is the same as that in Figure 3, the box #3 is different across these examples, mainly because the additional pattern is required to satisfy the individual reconstructability requirement of the subgraph mining problem. Finally, the subgraphs explored by the pattern matching engine are aggregated based on each input pattern in the pattern matching query, allowing the final results to be generated for each input pattern.

```
Count((π,1),
  Union(P₁,
    Union(P₂,P₃)))
```

(Quasi-Clique Mining)

```
Union(Count((π₁,1),P₁),
  Union(Count((π₂,1),P₂),
    Count((π₃,1),P₃)))
```

(Subgraph Mining)

Fig. 5. Queries for Quasi-Clique Mining and Subgraph mining.

## 2.2 Goals

In this work, we build an optimizing framework that enables evaluating pattern matching queries addressing graph mining problems. This framework supports optimizing both single and batched queries, supports a variety of notions of reconstructability, and enables user-defined optimizations that can work across a range of query and reconstructability requirements. In relation to our previous examples, the green boxes in Figures 3 and 4 delineate the scope of our proposed framework.

## 2.3 Our Approach

We solve this with our new Graph query E-graph-based Optimizer, Geo. Our high-level approach is to express these queries in a language well-suited for rewrite-based optimizations, and then use egg [Willsey et al. 2021] to efficiently traverse the query equivalences.

*2.3.1 Language.* Our language addresses graph mining problems via pattern matching queries. Our queries consist of a set of patterns to match, and postprocessing computations of those patterns. These two needs are captured by the two constructs: `Pattern` and `Count`. The `Pattern(P,F)` construct captures the exact subgraph to extract from the data graph using the provided pattern `P`, and a filter `F`. The `Count` construct is more complex. In short, the `Count(Π,Q)` construct distributes the results of the internal query `Q` across a variety of results. This reconstruction path Π encodes a mapping from problem domains $\pi$ to multiplicative factors. This enables sharing between batched calls – two queries may use the same subquery, despite using such a result in different ways. Figure 5 shows the queries for Quasi-Clique Mining and Subgraph Mining written in our language.

(1) *Quasi-Clique Mining.* The quasi-clique mining query involves summing the counts of the three viable quasi-cliques patterns – $P_1$, $P_2$, and $P_3$. These sub-results are then aggregated into a single final result identified by $\pi$.
(2) *Subgraph Mining*. Recall the subgraph mining problem. The subgraph mining is a union of three sub-queries, simply identifying the results for $P_1$, $P_2$, and $P_3$ and returning all of them. Because of their distinct identifiers $\pi_1$, $\pi_2$, and $\pi_3$, these results are *not* aggregated.

Note that currently these queries simply run three problems in isolation. There is essentially no postprocessing required – Geo simply asks the pattern mining engine to find the counts of each pattern, and then either adds them together, or keeps them separate. This is in fact desired – we want users to be able to maintain the abstraction that their pattern matching queries simply run independently. However, our optimizer should be able to break these queries into equivalent, faster queries that enable sharing of subquery results.

*2.3.2 Optimizing Pattern Queries.* Generally, it is much more efficient to run pattern matching queries that require matching (1) fewer patterns, (2) smaller patterns, and (3) patterns that are easier to match for pattern matching engines. However, this space of possible query optimizations is complex. These query optimizations involve theoretical insights from graph mining experts and a complex optimization problem. In this work, we separate these two problems: domain experts identify candidate rewrites, and Geo processes these rewrites with egg [Willsey et al. 2021].

However, care must be taken when applying these rewrites. Consider the query for counting two patterns: $p_a$ and $p_b$. Consider the two equivalence rewrites where the count of `Pattern(`$p_a$`)` is known to be equivalent to the number of Pattern($p_c$) subtracted by the number of Pattern($p_d$); as well as the equivalence where the count of Pattern($p_b$) is known to be equivalent to the number of Pattern($p_e$) plus by the number of Pattern($p_d$). Naïvely applying these equivalences will actually return incorrect results: the added Pattern($p_d$) and the subtracted Pattern($p_d$) cancel out. But this now removes the information required for reconstructing the original Pattern($p_a$) and Pattern($p_b$) results, breaking reconstructability. Typically, these optimizations are safely performed by keeping them isolated, then performing common subexpression elimination. However, common subexpression elimination in e-graphs is quite complex, requiring complex reasoning outside the native capabilities of e-graphs [Cao et al. 2023]. Fortunately, because of the $\Pi$ constructs in our language, we embed knowledge of how the Pattern($p_b$) construct is used to reconstruct the query results for $p_a$ and $p_b$, despite the rewrite. Furthermore, we only remove such queries when the result is not needed for every query in the batch. This is why Figure 3 requires matching only 2 patterns, whereas Figure 4 requires matching 3.

Moreover, there is some trickiness in integrating rewrite rules into Geo. In particular, there are multiple equivalences that operate on the patterns themselves. For example, the 4-cycle graph $P_4$ can be written as $a - b - c - d - a$ or as $b - c - d - a - b$. These *graph automorphisms* are not well-suited to a rewrite-based theory. Simply adding a whole bunch of rewrite rules that transform graphs into equivalent forms is not a good option, the search space becomes unreasonably large. To address this, we focus on only the patterns that have been put into canonical form. In essence, we replace each pattern in the query by a canonical one, where any two equivalent patterns get mapped to the same canonical form. This canonicalization is performed by Bliss [Junttila and Kaski 2007], a graph canonicalizer that we have used for pattern canonicalization. In essence, when provided a query, we immediately canonicalize that query to contain only canonical patterns.

However, performing these canonicalizations naïvely can cause issues: If we canonicalize $P$ to $P'$, but have a rewrite rule $P \rightarrow P''$, that rewrite rule no longer fires. This space even more tricky because performing these canonicalizations can not ensure that every term in our e-graph is fully canonicalized, due to subterms not necessarily being in canonical representations. Even though we canonicalize the full pattern, the individual subcomponents of the pattern may not be canonicalized.

To address this, we require that all rewrite rules (provided either by us or by rewrite developers) must *respect* the canonicalization functions. In essence, if a rule applies to some term, it *must also* apply to the canonicalized form of that term. The transformations that apply to canonicalized queries must also output the canonicalized form of the original rewrite result.

With canonicalization and reconstruction paths together, our solution transforms pattern matching queries to more efficient forms, without missing opportunities for complex interactions between different rewrite rules. This allows users to simply express their pattern matching queries and generate cost-optimized equivalent pattern queries. Furthermore, domain experts can simply express their own rewrite rules to enrich the set in Geo without worrying about how they would interact with existing rules as Geo automatically explores the space of possible applications of rewrites.

## 3 Pattern Matching Query Language: Syntax and Semantics

Our pattern matching query language aims to support three core features. Our language must support batching: multiple pattern mining problems must be able to be expressed in a single query, and subresults should be able to be shared across the batched queries. Our language must support reconstructability: our semantics must be rich enough that they can describe results for problems that look very distinct from the queries run. Lastly, our language must be amenable to optimization: the language should be relatively simple to make a search-based optimizer tractable.

Figure 6 shows the grammar of our language. Queries are either Unions, Counts, or Patterns.

$$
\begin{array}{lrcl}
\text{Queries} & Q & ::= & \texttt{Union}(Q_1, Q_2) \\
 & & \mid & \texttt{Count}(\Pi, Q) \\
 & & \mid & \texttt{Pattern}(P, F) \\
\text{Patterns} & P & ::= & \texttt{Edge}(v_1, v_2)\ P \\
 & & \mid & \texttt{Anti-Edge}(v_1, v_2)\ P \\
 & & \mid & \cdot \\
\text{Reconstruction Path} & \Pi & ::= & \Pi + \Pi \\
 & & \mid & (\pi, x)
\end{array}
$$

Fig. 6. Formal grammar of the pattern matching query language. The meta-variable $v$ ranges over an alphabet for vertex variables $\Sigma_v$. The meta-variable $\pi$ ranges over a set of provenances $\Sigma_\pi$. The meta-variable $x$ ranges over real numbers. The meta-variable $F$ denotes a function from patterns matches to boolean values.

`Union` queries are simple combinations of sub-queries. Unions are the core construct that enable batching the identification of pattern matches.

`Count` queries perform dual purposes: they *scale* and they *distribute*. When optimizing PMQs (Pattern Matching Queries), shared subqueries must be scaled and distributed to the original queries that spawned them. The reconstruction rath $\Pi$ describes the mapping from original query to multiplicative factor.

`Pattern` queries perform the core search. Pattern queries consist of a pattern $P$ and a filter $F$. The pattern matching engines can perform the core identification and filtering involved in $P$. If such an F is omitted, there is no filtering of the patterns matched.

The notion of reconstruction paths are core to our language, though the terms themselves are relatively simple. The provenances are essentially simple identifiers describing the individual components of a batched query. The provenance of 1 is a sentinel value identifying that a query result should be used for every individual component in the batch. This unital provenance is then refined to apply to individual components via applications of the `Count` rule.

$$
\begin{array}{rcl}
[\![\texttt{Union}(Q_1, Q_2)]\!]_G & = & \pi \mapsto [\![Q_1]\!]_G(\pi) + [\![Q_2]\!]_G(\pi) \\
[\![\texttt{Count}(\Pi, Q)]\!]_G & = & \pi \mapsto [\![\Pi]\!](\pi_1) \times [\![Q]\!]_G(\pi_2) \text{ where } \pi_1\pi_2 = \pi \\
[\![\texttt{Pattern}(P, F)]\!]_G & = & 1 \mapsto |\{\sigma \mid [\![P]\!](G, \sigma) \land F(\sigma)\}| \\
 & & \pi \mapsto 0 \text{ where } \pi \neq 1 \\
 & & \\
[\![\texttt{Edge}(v_1, v_2)\ P]\!] & = & (G, \sigma) \mapsto (\sigma(v_1), \sigma(v_2)) \in G.edges \land [\![P]\!](\sigma) \\
[\![\texttt{Anti-Edge}(v_1, v_2)\ P]\!] & = & (G, \sigma) \mapsto (\sigma(v_1), \sigma(v_2)) \notin G.edges \land [\![P]\!](\sigma) \\
[\![.]\!] & = & (G, \sigma) \mapsto \mathit{true} \\
 & & \\
[\![\Pi_1 + \Pi_2]\!] & = & \pi \mapsto [\![\Pi_1]\!](\pi) + [\![\Pi_2]\!](\pi) \\
[\![(\pi, x)]\!] & = & \pi' \mapsto x \text{ when } \exists \pi'' \text{ such that } \pi' = \pi\pi'' \\
 & & \pi' \mapsto 0 \text{ otherwise}
\end{array}
$$

Fig. 7. Semantics of the graph pattern mining language. Given a concrete data graph $G$, query semantics provide a mapping from provenance identifiers $\pi$ to real numbers. Given an assignment of vertex variables to vertices, patterns provide a mapping from such an assignment to bool. Reconstruction paths provide a mapping from provenances to multipliers.

Our language has three semantics: one for queries, one for patterns, and one for reconstruction paths, shown in Figure 7. Given a provenance identifier $\pi$, the semantics of a query $Q$ on graph $G$ ($[\![Q]\!]_G$) outputs a number for that provenance identifier. This approach enables reconstrucability: given the original query identifier $\pi$, we can reconstruct the original results by applying such a $\pi$ to the reconstruction path extracted from the semantics. The `Union` query, given an identifier $\pi$ simply sums the results of the sub-queries. The `Count` query distributes the results of subqueries according to the provided reconstruction path $\Pi$. Finally, the `Pattern` query actually evaluates the pattern using a pattern matching engine, and applies a further filter $F$ to the pattern matches.

Patterns run via *pattern matching engines.* A pattern candidate $\sigma$ for a graph $G$ is a mapping of vertex identifiers to vertices in a graph. This pattern candidate is a match for a pattern $P$ if it satisfies the `Edge` and `Anti-Edge` constraints described by the pattern semantics $[\![P]\!]$. These semantics are used in the `Pattern` combinator, which counts the number of matches (and assigns it a provenance of 1, to be allocated to provenances via the `Count` combinator). These provenances satisfy the idempotent semiring axioms, including the rules: $1\pi = \pi 1 = \pi$ and $\pi\pi = \pi$. These rules ensure that the unital provenance can be specialized to any arbitrary provenance, and that specializing a query to the same provenance a second time has no effect.

Finally, the semantics of a reconstruction path $[\![\Pi]\!]$ simply maps provenance variables to concrete numbers, describing the impact of individual patterns on various reconstruction results.

Through this language, we provide a relatively simple syntax that enables (1) batching, (2) result sharing, and (3) reconstructability across rewrites.

*Example: Subgraph Mining.* Recall our original subgraph mining query:

$$\texttt{Union(Count((}\pi_1\texttt{,1),}P_1\texttt{), Union(Count((}\pi_2\texttt{,1),}P_2\texttt{), Count((}\pi_3\texttt{,1),}P_3\texttt{)))}$$

The semantics of this query is relatively simple to follow: find the total number of instances of $P_1$, $P_2$, and $P_3$; and return each of those query results independently.

With these flexible semantics, reconstructability simply becomes semantic equivalence. As long as our rewrites maintain semantics, the results returned by running the optimized query are the same as those from running the original query.

Consider the semantics of the optimized subgraph mining query:

$$\begin{aligned}&\texttt{Union(Count((}\pi_1\texttt{,1)+(}\pi_2\texttt{,3)+(}\pi_3\texttt{,-6),}P_1\texttt{),}\\&\quad\texttt{Union(Count((}\pi_2\texttt{,1),}P_4\texttt{),}\\&\qquad\texttt{Count((}\pi_2\texttt{,-1)+(}\pi_3\texttt{,1),}P_5\texttt{)))}\end{aligned}$$

The semantics of this query is more complex. The computation for $P_1$ is unchanged – it is directly computed. However, the computations for $P_2$ and $P_3$ are substantially altered, involving two different queries $P_4$ and $P_5$. By *batching* these queries together, we find 3 *shared substructures* that can be used to ~~re~~compute the original 3 patterns, and matching these smaller patterns is faster than matching the originals. Then, the semantics of this new query combines the counts of these substructures in unique ways to *reconstruct* the results of the original query.

## 4 Rewrite-Based Pattern Matching Query Optimizations

Running the direct translation of a user's graph mining problem is often rather inefficient, as the patterns being matched upon are often large and inefficient for pattern matching engines. Thus, optimization is key for solving complex graph mining problems.

Nearly all of the computation involved in evaluating a query comes from the time spent evaluating pattern matching queries. Thus, similar to existing works [Jamshidi et al. 2023], we simply ignore the cost of all other operations – the cost of a query is just the cost of all patterns matched.

*Definition 4.1.* We can lift a cost function $C$ : `Pattern` $\rightarrow \mathbb{R}$ defined on patterns to a cost on queries, defined as $C(Q) = \sum_{p \in P} C(p)$

We aim to optimize our queries to minimize this cost. We want this optimization algorithm to be easily extensible, enabling graph experts to easily bring their own equivalences, and GEO should be able to integrate them into its algorithm.

To this end, our algorithm operates over rewrites. As long as graph experts can provide their rewrites in a specific form, GEO will integrate them into its algorithm, and traverse those rewrites in the process of identifying a low-cost query.

In this section, we begin with a brief preliminaries subsection introducing notation and formalizing rewrites, before proceeding in describing the role of rewrites, and restrictions we place on them, in our full optimization algorithm.

### 4.1 Rewrite Preliminaries

The primary abstraction we work over is of *terms* over an arbitrary ranked alphabet. Terms correspond to pattern matching query expressions, and the ranked alphabet corresponds to the PMQ grammar syntax.

*Definition 4.2.* A *ranked alphabet* $\Sigma = (S, rank)$ is a pair of a set $S$ and a function $rank : S \to \mathbb{N}$. By convention, if $c \in S$, we can use the shorthand $c \in \Sigma$. Furthermore, $\Sigma_i$ refers to the set $\{c | c \in S \wedge rank(c) = i\}$.

*Definition 4.3.* Given a ranked alphabet $\Sigma$, the set of terms $\mathcal{T}(\Sigma)$ is the smallest set that satisfies if $c \in \Sigma_n$ and $t_i \in \mathcal{T}(\Sigma)$ then $c(t_0, \ldots, t_n) \in \mathcal{T}(\Sigma)$.

User-defined rewrite rules are introduced as rewrites. These rewrites consist of two components: a *matcher* and a *transformation function*. The matcher[2] is intuitively a term with holes, and it identifies which subexpressions can be rewritten via unifying a variable substitution. The transformation function describes what the matched subexpression is rewritten to.

*Definition 4.4.* Given an alphabet $\Sigma$, the set of matchers $\mathcal{M}(\Sigma)$ is defined as $\mathcal{T}(\Sigma \cup \mathcal{X})$ where $\mathcal{X}$ is a distinct infinite set of 0-ary *matcher variables*.

*Definition 4.5.* A substitution $\sigma : \mathcal{X} \to \mathcal{T}(\Sigma)$ is a mapping of matcher variables to terms. Given a matcher $m \in \mathcal{M}(\Sigma)$, we inductively define $m[\sigma]$ as $x[\sigma] = \sigma(x)$ when $x \in \mathcal{X}$ and $c(t_1, \ldots, t_n)[\sigma] = c(t_1[\sigma], \ldots, t_n[\sigma])$ when $c \in \Sigma$.

*Definition 4.6.* In a ranked alphabet $\Sigma$, a *rewrite rule* $r$ is a pair $r = (m, f)$ where $m \in \mathcal{M}(\Sigma)$ is a matcher and $f : (\mathcal{X} \to \mathcal{T}(\Sigma)) \to \mathcal{T}(\Sigma))$ is a mapping from substitutions to terms such that if $m[\sigma] = m[\sigma']$ then $f(\sigma) = f(\sigma')$. These rewrite rules are often written as $m_1 \to m_2$ (where the variables used in $m_2$ are a subset of those in $m_1$) to denote that the substitution is $m_2[\sigma]$.

Finally, we show how to rewrite according to a set of rewrites $R$. One can rewrite a term if either that term is matched by a matcher in $R$, or if some subterm is matched by a matcher in $R$.

*Definition 4.7.* Let $R$ be a set of rewrites and $t_1$ and $t_2$ terms. We say that $t_1 \to_R t_2$ if either:

(1) $(m, f) \in R$ and there exists some $\sigma$ such that $m[\sigma] = t_1$ and $f(\sigma) = t_2$
(2) If $t_1 = c(u_1, \ldots, u_n)$ and $t_2 = c(u'_1, \ldots, u'_n)$ and $u_i \to_R u'_i$ or $u_i = u'_i$

1. $\texttt{Union}(Q_1, Q_2) \leftrightarrow \texttt{Union}(Q_2, Q_1)$
2. $\texttt{Union}(Q_1, \texttt{Union}(Q_2, Q_3)) \leftrightarrow \texttt{Union}(\texttt{Union}(Q_1, Q_2), Q_3)$
3. $\texttt{Count}(\Pi, \texttt{Union}(Q_1, Q_2)) \to \texttt{Union}(\texttt{Count}(\Pi, Q_1), \texttt{Count}(\Pi, Q_2))$
4. $\texttt{Count}((\pi_1, n_1), \texttt{Count}((\pi_2, n_2), Q)) \to \texttt{Count}((\pi_1\pi_2, n_1 \times n_2), Q)$
5. $\texttt{Union}(\texttt{Count}((\pi, n_1), \texttt{Pattern}(P, F)), \texttt{Count}((\pi, n_2), \texttt{Pattern}(P, F)))$
   $\to \texttt{Count}((\pi, n_1 + n_2), \texttt{Pattern}(P, F))$
6. $\texttt{Union}(\texttt{Count}(\Pi_1, \texttt{Pattern}(P, F)), \texttt{Count}(\Pi_2, \texttt{Pattern}(P, F)))$
   $\to \texttt{Count}(\Pi_1 + \Pi_2, \texttt{Pattern}(P, F))$
7. $\texttt{Pattern}(P_1, F) \to \texttt{Pattern}(P_2, F)$ if $P_1 \equiv_{Graph} P_2$

Fig. 8. Built-In Rewrite Rules. Rules marked with $\to$ indicate transformations that can be applied in one direction, from left to right. Rules marked with $\leftrightarrow$ can be applied in either direction.

[2] We use the word “matchers” instead of the more common term of “patterns” to avoid overloading the term.

### 4.2 Built-In Rewrites

Our system includes several built-in rewrite rules, shown in Figure 8. These rules ensure both the correctness and optimization of query transformations, enabling effective algebraic manipulations over graph queries.

*Commutativity and Associativity (Rules 1 and 2).* These two rewrite rules define the commutativity and associativity of `Union`, ensuring that the order of pattern unions does not affect their combined representation.

*Distributivity of* `Count` *Over* `Union` *(Rule 3).* : This rule expresses the distributive property of `Count` with respect to `Union`, enabling individual counts of pattern components in a unioned structure.

*Nesting of Counts (Rule 4).* : This rule handles multiplicative nesting of `Count` operations, where nested counts are combined by multiplying their coefficients.

*Pattern Combination with Reconstruction Paths (Rules 5 and 6).* : The last two rules merge patterns with identical structures but differing reconstruction paths:

- If the patterns are the same and are used for reconstructing the same query result, we add their coefficients.
- If the patterns are identical but are assigned different reconstruction paths, we sum those paths.

These rules ensure that when data gets merged, it is done in a way that maintains reconstructability.

*Pattern Equivalence (Rule 7).* This rule allows us to treat semantically equivalent patterns as identical, effectively eliminating redundant representations. In the context of graphs, equivalence leverages the concept of *isomorphism*, which establishes when two graphs represent the same structure.

Unfortunately, this Rule 7 is both important and problematic, and so we find an alternative way of dealing with those equivalences. Without Rule 7, GEO fails at enabling a number of optimizations. It would not share subquery results unless those subqueries were written, and automatically rewritten by various rewrites, in the exact same way. Furthermore, if the queries operate on specific patterns (like the ESCAPE rules), then those rules only fire if the patterns are written in that exact way.

However, applying Rule 7 naïvely creates an unwieldy search space. Every pattern in every rule can be rewritten to any *automorphism* of that pattern – any pattern that is equivalent under renaming. Instead, we perform a lightweight canonicalization over the patterns. Then, as long as all rewrites “play nicely” with canonicalizations, we can completely elide Rule 7.

### 4.3 Taming the Search Space with Canonicalizations

Canonicalizations are functions that pick representatives from equivalence classes. A canonicalized term is one that is the representative of the equivalence class it inhabits. We additionally require that if two terms are built from subterms that are in the same canonicalization-induced equivalence class, they should canonicalize to the same output.

*Definition 4.8.* A canonicalization $\varphi$ is a function $\varphi : \mathcal{T}(\Sigma) \to \mathcal{T}(\Sigma)$ such that $\varphi \circ \varphi = \varphi$ and $\varphi(t_i) = \varphi(t'_i)$ implies that $\varphi(c(t_1, \ldots, t_n)) = \varphi(c(t'_1, \ldots, t'_n))$. Given a canonicalization function $\varphi$, a term $t$ is canonicalized according to $\varphi$ if $\varphi(t) = t$.

However, simply canonicalizing our input pattern is not enough. We require restrictions over cost functions and rewrite rules to ensure that they agree with the canonicalized functions.

*Definition 4.9.* A set of rewrites $R$ respects a canonicalization $\varphi$ if for all terms $t$ and $t'$, $t \to_R t'$ implies that $\varphi(t) \to_R \varphi(t')$

*Definition 4.10.* A cost function $C : \mathcal{T}(\Sigma) \to \mathbb{R}$ respects a canonicalization $\varphi$ if, for all $t$, $C(\varphi(t)) \leq C(t)$

These restrictions on rewrite rules and cost functions are essential because they allow our approach to work exclusively with terms in canonical form. If a rewrite rule did not respect a canonicalization, then our algorithm may not process all the possible rewrites – as some rewrites may require a specific form. Similarly, if a cost function does not respect canonicalization, it could assign different costs to equivalent terms based on their representation, misleading the optimization process. This inconsistency would imply that two identical patterns (in terms of structure) could appear to have different costs, causing the algorithm to overlook optimal solutions.

We aim to have our pattern matching query language canonicalize the patterns of the query, and leave the other components unchanged. We define this canonicalization function as $\varphi_{PMQ}$.

*Definition 4.11.* If $t$ is a pattern, then $\varphi_{PMQ}(t) = \varphi_{Graph}(t)$, where $\varphi_{Graph}$ is a graph canonicalization algorithm. If $t$ is not a pattern, and $t = c(t_1, \ldots, t_n)$, then $\varphi_{PMQ}(t) = c(\varphi_{PMQ}(t_1), \ldots, \varphi_{PMQ}(t_n))$.

Critically, this canonicalization function is sound – canonicalizing our patterns does not change the semantics.

Theorem 4.12. *Let $Q$ be a graph query. For all graphs $G$, $[\![Q]\!]_G = [\![\varphi_{PMQ}(Q)]\!]_G$*

Thus, we require an additional restriction on the user. If they wish to provide additional rewrites, then they must manually ensure that their rewrites respect canonicalization. However, we have found this is not too difficult in practice. For our canonicalization function, we have found that if the matchers of rewrite rules treat patterns *opaquely*, it is easy to make a rewrite rule respect canonicalization.

*Definition 4.13.* A matcher $m$ treats patterns opaquely if it conforms to the grammar in Figure 7, augmented to allow matcher variables in all productions except those of patterns.

Essentially, any time a variable matches a pattern, the full pattern must be matched. This means that matchers like the single variable $p$ or the matcher `Edge(`$v_1$`,`$v_2$`)` $p$ would not be permitted, but matchers like `Pattern(`$p$`)` or `Pattern(Edge(`$v_1$`,`$v_2$`))` would. Furthermore, every transformation function must output a canonicalized term, equivalent to the output of running the transformation function on a substitution with variables mapping to equivalent queries.

Theorem 4.14. *Let $(m, f)$ be a rewrite rule. If $m$ treats patterns opaquely, then the rewrite rule $(\varphi_{PMQ}(m), \varphi_{PMV} \circ f \circ \varphi_{PMQ})$ respects $\varphi$.*

### 4.4 Adding Custom Rewrites

Custom rewrites are relatively simple to write. As an example, consider the rewrite reducing the match of a 3-cycle to the problem of wedges (i.e. three nodes connected by two edges). The number of 3-cycles is equivalent to a third of the number of wedges, minus a third of the number of wedges with an explicit anti-edge. The following is a rewrite that encodes this equivalence:

```
φPMQ(Pattern(Edge(v1,v2) Edge(v2,v3) Edge(v3,v1))) ->
φPMQ(Union(Count((1,1/3),Pattern(Edge(v1,v2) Edge(v2,v3))),
  Count((1,-1/3),Pattern(Edge(v1,v2) Edge(v2,v3) Anti-Edge(v1,v3)))))
```

Notably, we can ensure that this rewrite rule respects $\varphi_{PMQ}$ simply by applying it to both sides of the rule.

## 5 Optimization Algorithm

Given a set of input patterns $P$, a set of rewrite rules $R$ for transforming patterns into equivalent alternative forms, and a cost function $C$ estimating the computational expense of each pattern, our objective is to find an optimal final set of patterns that yields the same results as the initial input patterns while minimizing the associated cost.

**Algorithm 1** Pattern Optimization using E-Graphs

**Require:** Query $Q$, Set of rewrite rules $R$, Cost Function $C$, a set of stop conditions $S$
**Ensure:** Optimal solution with minimized pattern costs
1: $R' = (R \cup (B \setminus \text{Rule 7}))$
2: Initialize e-graph $E$ with Canonicalize($Q$)
3: **while** $\bigwedge_{s \in S} \neg s$ **do**
4: $\quad E = R'(E)$
5: **for** each pattern in $E$ **do**
6: $\quad$ Calculate cost of pattern using $C$
7: $E$.find(Canonicalize($Q$))

The solution must ensure two key requirements:

(1) The results derived from the alternative patterns must be transformable, via the rewrite rules, to match exactly the results that would be obtained from the original input patterns.
(2) The selected alternative patterns should minimize the estimated cost function, achieving optimal efficiency in the search process.

More formally, we define our optimization problem statement as follows:

**Optimization Problem Statement**: Given a pattern matching query $Q$, a set of rewrite rules $R$ comprised of built-in rewrite rules $B$ and additional custom rewrite rules $A$ that respect $\varphi_{GPM}$, and a cost function over queries $C$ that respects $\varphi_{GPM}$, find a query $Q'$ such that $Q \to_R^* Q'$ and $\forall Q''$, $Q \to_R^* Q'' \implies C(Q') \leq C(Q'')$.

Our algorithm optimizes the provided query by repeatedly applying rewrite rules. Because of its efficiency performing rewrite-based optimizations, we embed our rewrites in egg [Willsey et al. 2021], which automates the process of repeatedly applying rewrite rules, and identifying expressions made equivalent by them. These e-graphs support 3 primary operations: (1) we can initialize the e-graph with a term, (2) we can apply all the rewrite rules once (and ensure the resultant equivalence function is congruence closed), and (3) we can perform the *find* operation, which identifies a minimal cost term equivalent to the input term.

We begin by taking an input query $Q$ from the user, along with required components: transformation rules and a cost function. These components, along with built-in rewrite rules and stop conditions $S$, configure the system to dynamically generate and evaluate alternative patterns, facilitating flexible pattern optimization. Our algorithm is presented in Algorithm 1.

### 5.1 Steps in the Algorithm

*1. Input Initialization.* The algorithm starts with a user-provided query $Q$, which consists of a set of input patterns. First, each pattern in $Q$ is canonicalized. This step ensures that structurally identical patterns with different representations are stored only once in the e-graph. The e-graph is then initialized with the canonical form of $Q$.

*2. Rule Application.* The algorithm then enters a main loop where rewrite rules are recursively applied to the patterns in the e-graph. There are two types of rules:

- *Custom Rewrite Rules:* These rules transform a given canonicalized pattern into one or more alternative canonicalized patterns. The rule triggers when it matches the syntax of a pattern alongside its reconstruction path. Newly generated patterns are added to the e-graph. The e-graph changes accordingly to capture this new equivalence of patterns.

- *Built-in Rewrite Rules:* These rules apply algebraic transformations across patterns, allowing further simplification and optimization as new patterns are generated. Notably, we do not need to use Rule 7, as this rule is addressed via canonicalization.

*3. Stop Condition.* The loop of rule applications continues until one of the specified stopping conditions $s \in S$ is met. Upon reaching the stop condition, the e-graph is finalized.

*4. Cost Evaluation and Extraction.* With the e-graph complete, the system evaluates the cost of each pattern using the user-defined cost function. The extractor retrieves the optimal pattern set by selecting the minimal-cost path through the e-graph. The final result provides the optimal solution with minimized pattern costs.

### 5.2 Correctness

We have proven that our algorithm is correct. Given a set of user-provided rewrites that respect our canonicalization function, if our Algorithm achieves saturation, we are guaranteed to extract the minimal query reachable in some number of rewrites.

THEOREM 5.1. *Let $R$ respect $\varphi_{GPM}$ and $S = \{R'(E) = E\}$ and Algorithm 1 returns a query $Q'$. This $Q'$ is a minimal cost query that satisfies $Q \rightarrow^*_{R \cup A} Q'$.*

Notably, this theorem is true despite the fact we do not apply any automorphism rewrite rules, due to our choice of canonicalization function, and due to our assumptions about $R$ and $C$. The proof of this theorem is available in Appendix B.

## 6 Implementation

In this section, we discuss the practical aspects involved in implementing GEO.

*Rewrite Rules and Pattern Generation.* GEO is instantiated with rewrite rules, which are specified in terms of patterns. These rewrite rules can be defined statically for each input pattern or generated dynamically using well-defined formulas. For our evaluation, ESCAPE rules are specified statically as a set of 15 rules, whereas SUBGRAPH MORPHING rules are generated dynamically using a generalized formula.

*Cost Function.* The cost of a query is derived from the costs of the patterns it comprises. Generating efficient equivalences requires a cost model for patterns that estimates the time taken to mine them. We model the performance of mining different patterns in a dataset using the relative cost model designed in [Jamshidi et al. 2023], which assigns integer estimates for patterns indicating the relative time they would take to mine.

*Termination Criteria.* Rule application within the e-graph ceases when any one of the following stopping conditions is met: (1) *saturation*, meaning that the e-graph has captured all equivalent expressions; (2) a *time limit*, indicating the maximum allowable time for saturation; (3) an *iteration limit*, indicating the maximum number of saturation rounds; or (4) a *node limit*, which constrains the e-graph's size. GEO exposes these parameters.

## 7 Evaluation

Graph mining is computationally expensive as subgraph isomorphism is NP-complete. Hence, pattern matching queries on large graphs often takes hours to weeks depending on the size of graphs [Jamshidi et al. 2020]. GEO's ability to explore novel query equivalences helps reduce the execution time for such expensive pattern matching queries.

To demonstrate the effectiveness of GEO, we instantiated it with two sets of rewrite rules that are of different types. The first one is SUBGRAPH MORPHING [Jamshidi et al. 2023], which is a generic algebra over subgraph structures and can be invoked on any arbitrary pattern. The equivalences in Subgraph Morphing are primarily based on sub-pattern/super-pattern relationships across same-sized patterns, i.e., in terms of denser or sparser patterns with the same number of vertices as the query patterns. The second is a set of 15 custom rules from ESCAPE [Pinar et al. 2017] that are applied to specific size-4 and size-5 patterns. These ESCAPE rules provide equivalences in terms of pattern decomposition, i.e., in terms of smaller-sized patterns compared to the query patterns. Figure 9 summarizes the characteristics of the two sets of rewrite rules.

| | Subgraph Morphing | ESCAPE |
|---|---|---|
| **Generation Method** | Dynamic from generalized formula | Static, Pre-defined |
| **Equivalence** | Sub-pattern/Super-pattern relationship | Pattern Decomposition |
| **Optimization** | Cost-based Search | None |
| **Anti-Edge Support** | Yes | No |

Fig. 9. Characteristics of the two sets of rewrite rules, Subgraph Morphing and ESCAPE, used in Geo.

With this instantiation, our evaluation aims to answer the following research questions:

***RQ1.*** Are Geo's novel query equivalences effective in reducing the execution time of pattern matching queries? This question assesses Geo's ability to automatically explore nuanced interactions between different kinds of rewrite rules and identify efficient query equivalences.

***RQ2.*** How Geo's EmRec choices improve the performance for pattern matching queries with collective reconstructability? This question evaluates the ability of Geo to optimize batched queries with collective reconstructability and contrasts it with the results for equivalent cases but with individual reconstructability.

***RQ3.*** How well does canonicalization help Geo in identifying useful rewrite rules? This question assesses the need for canonicalization in Geo and its impact on efficiency.

### 7.1 Experimental Setup

*Environment.* All experiments were conducted on a server equipped with an Intel Xeon Gold 6242R processor running at 3.10 GHz, with 40 physical cores and 62GB of RAM. Geo was built with Egg [Willsey et al. 2021] version 0.9.5 and Bliss [Junttila and Kaski 2007] version 0.77.

*Configuration.* We set the following Egg configurations: a time limit of 1 minute per run in the e-graph, an iteration limit of 40, and a node limit of 100,000 in the e-graph data structure. Each experiment was run for a maximum of 10 minutes. For executions that exceeded this timeout, the best computed equivalence rules from those executions were considered.

*Patterns, Dataset, and Cost Modeling.* The set of patterns used in our experiments is shown in Figure 10. Excluding anti-edges, there are 2 unique patterns of size three, 6 of size four, and 21 of size five. We selected these patterns to ensure a diverse representation of structures, including those used in state-of-the-art evaluations [Dias et al. 2019; Jamshidi et al. 2023; Shi et al. 2020].

We evaluated using the Friendster data graph [Elseidy et al. 2014], which contains approximately 65M nodes and 1.8 billion edges. This data graph was used to inform the pattern cost model, similar to the relative nature of the cost model in Subgraph Morphing which estimates the relative cost of mining each pattern based on: (a) the query pattern structure; (b) the data graph; and (c) the specifics of the mining engine. Here, we are using the Peregrine mining engine [Jamshidi et al. 2020; Jamshidi and Vora 2021] also used as the primary engine in the Subgraph Morphing research.

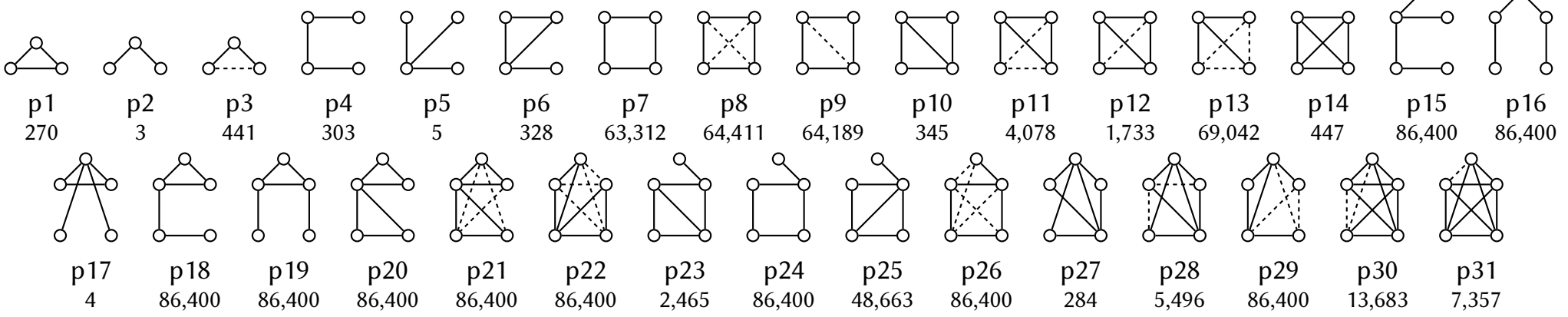

Fig. 10. Patterns used in evaluation along with their relative execution time (seconds) used for optimization.

Depending on the pattern complexities, the pattern matching queries can take a very long time on our Friendster data graph. We set a timeout of 1 day (86,400 seconds) for our pattern matching experiments, and hence several of our speedup numbers are conservative since baseline executions (, without Geo equivalences) often cross this timeout.

*Metric.* When evaluating the generated patterns, we run the generated pattern queries on the Friendster dataset, and measure the execution time in seconds.

### 7.2 Performance from Geo's Novel Equivalences (*RQ1*)

We analyze the performance of subgraph mining queries resulting from Geo compared to those from Subgraph Morphing and ESCAPE. We conduct two experiments. First, we evaluate single-pattern queries, representing the simplest form of query in a graph mining system but also being a difficult case for optimization since the space of equivalent queries here is limited. Next, we increase the complexity by testing batched-individual pattern queries, which contain multiple patterns with reconstructability requirements.

ESCAPE does not support batched queries, applying its rules independently to each pattern, without optimizing the rules according to a cost estimation. This often leads to missed optimization opportunities and can even increase the overall cost. In contrast, Subgraph Morphing supports batched queries, by naturally ensuring reconstructability. Unlike the baselines, Geo handles the individual reconstructability requirement using EmRec, and its combined use of different rewrite types leads to new and more effective optimizations.

*7.2.1 Singe Pattern Matching Queries.* Figure 11 shows the execution time of pattern matching queries resulting from Geo, ESCAPE and Subgraph Morphing. We observe that the optimized pattern matching queries generated by Geo consistently outperform or match those from ESCAPE and Subgraph Morphing. Compared to the original queries, Geo brings 1.3–320.0× speedup in execution time, excluding the one pattern where the original pattern itself is the best performing option and Geo also returns the same option.

| | **p11** | **p13** | **p15** | **p19** | **p20** | **p22** | **p24** | **p28** |
|---|---|---|---|---|---|---|---|---|
| Geo | **987** | **987** | **270** | **540** | **540** | **66,353** | **63,582** | 2,823 |
| Subgraph Morphing | 1,120 | 1,125 | 86,400 | 86,400 | 86,400 | 86,400 | 86,400 | 5,496 |
| ESCAPE | 4,078 | 69,042 | 329 | 615 | 615 | 86,400 | 63,657 | 5,496 |
| Original Query | 4,078 | 69,042 | 86,400 | 86,400 | 86,400 | 86,400 | 86,400 | 5,496 |
| Speedup | 4.13× | 69.95× | 320.00× | 160.00× | 160.00× | 1.30× | 1.36× | 1.94× |

Fig. 11. Execution time (seconds) of alternative pattern queries generated by Geo, ESCAPE, and Subgraph Morphing for various single input pattern queries. Speedups are relative to the original query.

| | **{p1, p3}** | **{p3, p4 }** | **{p6, p10, p13 }** | **{p11, p12 }** | **{p11, p15, p16 }** | **{p18, p19}** | **{p12, p23, p24}** | **{p20, p27}** | **{p30, p31}** |
|---|---|---|---|---|---|---|---|---|---|
| Geo | **273** | **273** | **987** | **987** | **64,569** | **810** | **64,299** | **810** | **6,388** |
| Subgraph Morphing | **273** | 745 | 1,125 | 1,120 | 173,920 | 172,800 | 89,657 | 86,684 | **6,388** |
| ESCAPE | 711 | 711 | 69,582 | 5,811 | 67,989 | 885 | 66,107 | 885 | 21,040 |
| Original Query | 711 | 745 | 69,716 | 5,811 | 176,878 | 172,800 | 90,598 | 86,684 | 21,040 |
| Speedup | 2.60× | 2.73× | 70.63× | 5.89× | 2.74× | 213.33× | 1.41× | 107.02× | 3.29× |

Fig. 12. Execution time (seconds) of alternative pattern queries generated by Geo, ESCAPE, and Subgraph Morphing for various batched-individual pattern queries. Speedups are relative to the original query.

| | GEO | SUBGRAPH MORPHING | ESCAPE |
|---|---|---|---|
| Pattern($p13$) | Union(Count(($\pi, -1$),Pattern($p1,F_1$)), Union(Count(($\pi, 2$),Pattern($p1,F_2$)), Count(($\pi, -4$),Pattern($p14$)))) | Union(Count(($\pi, 1$),Pattern($p5$)), Union(Count(($\pi, -1$),Pattern($p6$)), Union(Count(($\pi, 2$),Pattern($p10$)), Count(($\pi, -4$),Pattern($p14$))))) | Pattern($p13$) |
| **Time** | **987 sec** | **1,125 sec** | **69,042 sec** |

Fig. 13. Equivalences generated for $p13$ input by GEO, SUBGRAPH MORPHING and ESCAPE along with their run times. $F_1$ filters triangle nodes with external connections. $F_2$ filters triangle edges shared with other triangles.

| | GEO | SUBGRAPH MORPHING | ESCAPE |
|---|---|---|---|
| Pattern($p18$) | Union(Count(($\pi_{18}, 1$),Pattern($p1,F_3$)), Count(($\pi_{18}, -4$) + ($\pi_{19}, -2$), Pattern($p1,F_2$))) | Pattern($p18$) | Union(Count(($\pi_{18}, 1$),Pattern($p1,F_3$)), Count(($\pi_{18}, -4$),Pattern($p10$))) |
| Pattern($p19$) | Union(Count(($\pi_{19}, 1$),Pattern($p1,F_4$)), Count($\pi_{18}, -4$) + ($\pi_{19}, -2$), Pattern($p1,F_2$))) | Pattern($p19$) | Union(Count(($\pi_{19}, 1$),Pattern($p1,F_4$)), Count(($\pi_{19}, -2$),Pattern($p10$))) |
| **Time** | **810 sec** | **172,800 sec** | **885 sec** |

Fig. 14. Equivalences generated for $\{p18, p19\}$ input by GEO, SUBGRAPH MORPHING and ESCAPE along with their run times. $F_2$ filters triangle edges shared with other triangles, $F_3$ filters triangle nodes with external connections of at least degree 2, and $F_4$ filters edges where both endpoints have external connections.

| | GEO | SUBGRAPH MORPHING | ESCAPE |
|---|---|---|---|
| Pattern($p12$) | Union( Count(($\pi_{12}, 1$) + ($\pi_{24}, -2$),Pattern($p1,F_2$)), Count(($\pi_{12}, -6$) + ($\pi_{23}, -4$),Pattern($p14$))) | Union( Count(($\pi_{12}, 1$), Pattern($p10$)) Count(($\pi_{12}, -6$), Pattern($p14$))) | Pattern($p12$) |
| Pattern($p23$) | Union( Count(($\pi_{23}, 1$),Pattern($p1,F_7$)), Count(($\pi_{12}, -6$) + ($\pi_{23}, -4$),Pattern($p14$))) | Pattern($p23$) | Union( Count(($\pi_{23}, 1$),Pattern($p1,F_7$)), Count(($\pi_{23}, -4$),Pattern($p14$))) |
| Pattern($p24$) | Union( Count(($\pi_{24}, 1$),Pattern($p7,F_8$)), Count(($\pi_{12}, 1$) + ($\pi_{24}, -2$),Pattern($p1,F_2$))) | Pattern($p24$) | Union( Count(($\pi_{24}, 1$),Pattern($p7,F_8$)), Count(($\pi_{24}, -2$),Pattern($p10$))) |
| **Time** | **64,299 sec** | **89,657 sec** | **66,107 sec** |

Fig. 15. Equivalences generated for $\{p12, p23, p24\}$ input by GEO, SUBGRAPH MORPHING and ESCAPE along with their run times. $F_2$ filters triangle edges shared with other triangles, $F_7$ filters triangle nodes where two are connected in an external triangle and the third has an external connection, and $F_8$ filters 4-cycle nodes with external connections.

Moreover, using GEO achieves a speedup of up to 70× compared to ESCAPE and up to 320× compared to SUBGRAPH MORPHING. All cases where the execution time using GEO is lower than that of both ESCAPE and SUBGRAPH MORPHING indicate novel equivalences discovered by GEO. This shows that GEO identifies alternative patterns that neither baseline can achieve on its own, and effectively combines rewrite rules, improving performance by up to 99%.

*7.2.2 Batched Pattern Matching Queries.* Next, we evaluate batched-individual queries. Batched queries offer more opportunities for optimization, as they allow GEO to find common substructures across multiple patterns. As shown in Figure 12, GEO consistently provides significant improvements in execution time. Compared to the original query, equivalences from GEO result in 1.41–213.0× speedup in execution time. On the other hand, GEO achieves 1.03–70.5× speedup compared to ESCAPE and up to 213× speedup compared to SUBGRAPH MORPHING. For $\{p_1, p_3\}$, the rewrite rules from ESCAPE did not provide any benefits and so GEO simply ends up matching SUBGRAPH MORPHING's performance.

*7.2.3 Analyzing Benefits.* The improvement in execution time of novel equivalences generated by Geo is significant. We study three examples to showcase Geo's effectiveness.

First, Figure 13 shows the equivalences produced for the query pattern $p13$ where Geo's equivalences are 70× faster than that of the original query and ESCAPE, and 1.14× faster than Subgraph Morphing. Here, ESCAPE failed to reduce the cost, as it cannot handle those patterns. In contrast, Subgraph Morphing provides a solution that takes 1,125 seconds. However, Geo produces a novel equivalence with the lowest execution time of 987 seconds by combining various rewrite rules from both ESCAPE and Subgraph Morphing.

Second, Figure 14 shows the equivalences generated for the query pattern set $\{p18, p19\}$. In this case, Subgraph Morphing failed to lower the cost. However, by incorporating rules from ESCAPE together with Subgraph Morphing, Geo produces the solution with the lowest execution time of 810 seconds (ESCAPE's solution gives 9.3% slowdown compared to Geo).

Finally, Figure 15 shows the equivalences generated for the query pattern set $\{p12, p23, p24\}$. Both ESCAPE and Subgraph Morphing generate equivalences with lower execution times compared to the original query, but their combination in Geo yields further improvements, achieving a 1.41× speedup over the original query, which is 28.3% and 2.8% improvement over Subgraph Morphing and ESCAPE respectively.

**Result #1**: Geo's novel equivalences are effective in improving performance of pattern matching queries, with improvements over orders of magnitude in several cases. By exploring interactions among different kinds of equivalences across patterns, Geo uncovers new optimizations, not explored in prior research.

| | **{p2, p4}** | **{p8, p10}** | **{p10, p12, p19}** | **{p6, p11, p15}** | **{p20, p21}** |
|---|---|---|---|---|---|
| Geo Collective | 273 | 63,759 | 717 | 717 | 66,353 |
| Geo Individual | 273 | 64,029 | 987 | 987 | 86,940 |
| Improvement | — | 0.42 % | 27.36 % | 27.36 % | 23.68 % |
| Original Query | 306 | 64,756 | 88,478 | 90,807 | 172,800 |
| Speedup | 1.12× | 1.02× | 123.40× | 126.65× | 2.60× |

Fig. 16. Execution time (seconds) for batched-individual and batched-collective queries on Geo. Improvement is for Geo Collective over Geo Individual. Speedups are for Geo Collective relative to the original query.

### 7.3 Embedded Reconstructions for Collective Reconstructability (*RQ2*)

To study how different choices in EmRec affect Geo's optimization, we evaluate Geo on batched-collective pattern matching queries, where results of the entire query is of interest instead of the individual results for each pattern.

We capture the collective reconstructability requirement by assigning the same reconstruction path to all patterns in the input query. This same reconstruction path approach reveals optimizations beyond what any other tools can achieve, hence enabling Geo to outperform both ESCAPE and Subgraph Morphing. Moreover, Geo finds new optimizations that differ from those generated under individual reconstructability, where each pattern has its unique reconstruction path. By enabling cancellations of redundant patterns under the same reconstruction path, our method highlights a novel type of optimization that has not been explored in prior research.

Figure 16 shows the execution times for pattern queries generated by Geo with the two reconstruction path choices. As we can see, setting the same reconstruction path for collective reconstructability requirement improves execution time by up to 27% (resulting in a speedup of 1.02–126.00× over the original queries), while for the $\{p_2, p_4\}$ case it ends up matching performance of when unique reconstruction paths are set. Figure 17 shows the queries generated by Geo for a batched pattern matching query $\{p_6, p_{11}, p_{15}\}$ for the two EMREC choices. Using unique reconstruction paths for each pattern prevents GEO from detecting and eliminating redundant results. By setting the same reconstruction path, GEO identifies and removes redundant patterns, improving the performance by an additional 27.4%.

| GEO with Individual Reconstructability | GEO with Collective Reconstructability |
|---|---|
| For $p6$: `Count(`$(\pi_6, 1) + (\pi_{11}, 1) + (\pi_{15}, -2)$`,Pattern(`$p1, F_1$`))`<br><br>For $p11$: `Pattern(`$p11$`) = Union(Count(`$(\pi_{11}, 6)$`,Pattern(`$p14$`)),`<br>`Union(Count(`$(\pi_{11}, -4)$`,Pattern(`$p1, F_2$`)),`<br>`Count(`$(\pi_6, 1) + (\pi_{11}, 1) + (\pi_{15}, -2)$`,Pattern(`$p1, F_1$`))))`<br><br>For $p15$: `Pattern(`$p15$`) = Count(`$(\pi_6, 1) + (\pi_{11}, 1) + (\pi_{15}, -2)$`,Pattern(`$p1, F_1$`))` | For $\{p6, p11, p15\}$:<br>`Union(Count(`$(\pi, -4)$`,Pattern(`$p1, F_2$`)),`<br>`Count(`$(\pi, 12)$`,Pattern(`$p14$`)))` |
| **Execution Time = 987 sec** | **Execution Time = 717 sec** |

Fig. 17. Equivalences generated for $\{p6, p11, p15\}$ by GEO for individual and collective results along with their costs. $F_1$ filters triangle nodes with external connections of at least degree 2, and $F_2$ filters triangle edges shared with other triangles.

**Result #2**: GEO introduces a new form of optimization for batched-collective queries by setting the same reconstruction path, which further optimizes query costs. This results in optimized query rewrites that have not been explored in prior research.

### 7.4 Importance of Canonicalization (*RQ3*)

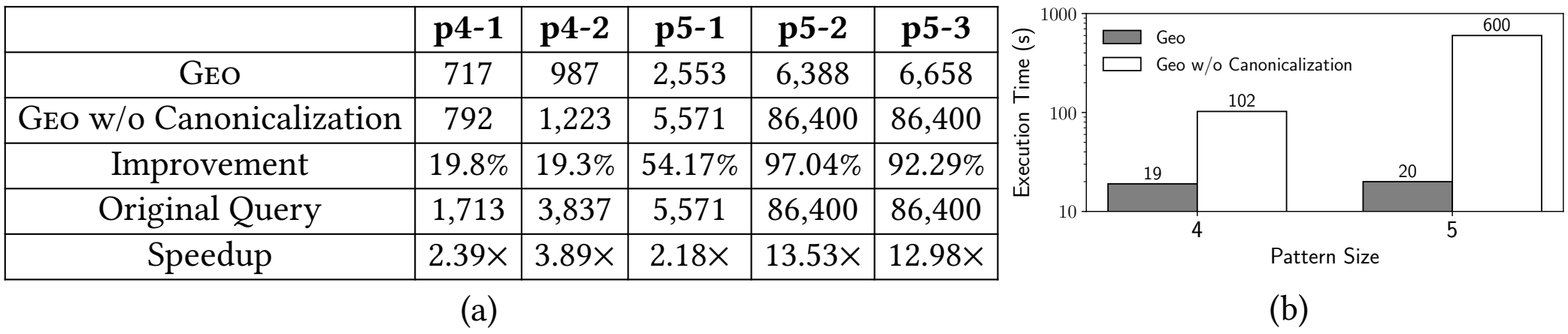

| | **p4-1** | **p4-2** | **p5-1** | **p5-2** | **p5-3** |
|---|---|---|---|---|---|
| GEO | 717 | 987 | 2,553 | 6,388 | 6,658 |
| GEO w/o Canonicalization | 792 | 1,223 | 5,571 | 86,400 | 86,400 |
| Improvement | 19.8% | 19.3% | 54.17% | 97.04% | 92.29% |
| Original Query | 1,713 | 3,837 | 5,571 | 86,400 | 86,400 |
| Speedup | 2.39× | 3.89× | 2.18× | 13.53× | 12.98× |

(a)

(b)

Fig. 18. (a) Execution time (seconds) of alternative patterns generated for GEO with and without canonicalization. Improvement is for GEO when canonicalization is enabled. Speedups are for GEO relative to the original query. (b) Execution times (seconds) of GEO (total time taken to generate alternative patterns) with and without canonicalization.

We study the effectiveness of canonicalization by disabling it in GEO and comparing the performance of the generated equivalences. Without canonicalization, duplicate patterns emerge in the e-graph even though the patterns are semantically equivalent, as they differ syntactically. Such redundancy would limit GEO's ability to identify opportunities to lower rewrite costs.

Figure 18(a) compares the execution times of generated equivalences when canonicalization is disabled in GEO versus when it is enabled. Since both the number of vertices in the pattern and the number of anti-edges affect optimization opportunities, we varied both of these parameters as part of the query patterns on the x-axis; hence, each input *px-y* represents a pattern with $x$ vertices and $y$ anti-edges, where the remaining edges are fully connected.

As seen, when canonicalization is disabled, GEO's optimization is limited as pattern complexity increases. As the number of vertices or anti-edges increases in patterns, the rewrite rules without canonicalization have lower performance compared to the equivalences generated by the canonicalized version. Benefits of canonicalization result in 2.18–12.98× overall speedup as it identifies and eliminates redundant patterns during e-graph saturation.

| | $APM_1$ | $APM_2$ | $APM_3$ | $APM_4$ | $APM_5$ |
|---|---|---|---|---|---|
| **Approximate Pattern Mining Query** | | | | | |
| **Patterns in Translated Batched Query** | | | | | |

Fig. 19. Approximate pattern mining queries with edit distance $k \leq 1$, and the set of patterns in their translated translated batched pattern matching queries.

Figure 18(b) shows Geo's execution times for these experiments. When canonicalization is disabled, Geo takes much more time to generate equivalences than when canonicalization is enabled. This is because without canonicalization Geo performs redundant equivalence explorations on identical patterns simply as they differ in syntax. This demonstrates the need for canonicalization for pattern matching query performance and Geo's efficiency.

**Result #3**: Canonicalization is important for recognizing and eliminating identical patterns in the e-graph. As pattern size and complexity increases, canonicalization helps Geo generate efficient equivalences and reduces Geo's execution time.

## 8 Case Studies

We conduct two case studies to evaluate the effectiveness of Geo on graph mining problems.

### 8.1 Approximate Pattern Mining

Approximate pattern mining aims to explore subgraphs that are structurally similar to a specific pattern. This is useful for various applications; for example, subgraphs that need further inspection in genomics pipelines [Paten et al. 2017], similar graphlets for computing social graph similarity, and exploring frequent patterns for fraud detection systems [Iyer et al. 2018]. The approximation in pattern structures is captured using *edit distance* which allows subgraphs to differ by a few edges.

| Approximate Mining Query | Optimized Query Size | Original Query Time | Geo Optimized Time | Time Reduction |
|---|---|---|---|---|
| $APM_1$ | 7 | 159,520 | **64,569** | 60% |
| $APM_2$ | 4 | 345,600 | 345,600 | — |
| $APM_3$ | 3 | 259,200 | 259,200 | — |
| $APM_4$ | 10 | 435,835 | **264,302** | 40% |
| $APM_5$ | 5 | 270,397 | **97,261** | 64% |

Fig. 20. Execution Time (seconds) for approximate pattern mining with collective reconstructibility.

| Approximate Mining Query | Optimized Query Size | Original Query Time | Geo Optimized Time | Time Reduction |
|---|---|---|---|---|
| $APM_1$ | 8 | 159,520 | **65,346** | 59% |
| $APM_2$ | 4 | 345,600 | 345,600 | — |
| $APM_3$ | 3 | 259,200 | 259,200 | — |
| $APM_4$ | 10 | 435,835 | **303,221** | 30% |
| $APM_5$ | 6 | 270,397 | **101,096** | 62% |

Fig. 21. Execution Time (seconds) for approximate pattern mining with individual reconstructibility.

We perform approximate pattern mining as defined in [Reza et al. 2020], which involves finding all patterns that are within a given *edit distance* $k$ from an input pattern $p$. In this context, edit distance is defined by the number of edge deletions required to make a pattern isomorphic to the input pattern. Hence, given an input pattern $p$, and an edit distance $k$, the approximate pattern mining problem translates into a batched pattern matching query containing patterns $\{p_1, p_2, ..\}$ such that each $p_i$ could become isomorphic to $p$ by adding at most $k$ edges. Figure 19 shows the input patterns used for approximate matching and the corresponding batched queries. Each batched query contains all patterns that are within an edit distance of $k \leq 1$ from the respective input pattern. For these queries, we study both the cases of reconstructability, as discussed below.

| Quasi-Clique Mining Query | $QCM_1$ | | $QCM_2$ | | $QCM_3$ | | $QCM_4$ | |
|---|---|---|---|---|---|---|---|---|
| | # vertices = 4 | $\gamma = 0.8$ | # vertices = 4 | $\gamma = 0.5$ | # vertices = 5 | $\gamma = 0.8$ | # vertices = 5 | $\gamma = 0.5$ |
| Patterns in Translated Batched Query | | | | | | | | |

Fig. 22. Quasi-clique mining queries and the set of patterns in translated batched pattern matching queries.

| Quasi-Clique Query | Optimized Query Size | Input Query Runtime | Optimized Runtime | Runtime Reduction |
|---|---|---|---|---|
| $QCM_1$ | 2 | 66,591 | **63,759** | 4% |
| $QCM_2$ | 7 | 226,111 | **64,569** | 71% |
| $QCM_3$ | 3 | 95,281 | **58,142** | 39% |
| $QCM_4$ | 10 | 700,081 | **321,717** | 54% |

Fig. 23. Runtime (seconds) for quasi-clique mining.

*8.1.1 Collective Reconstructability.* In this case, the results for all the patterns in the batched query are aggregated into a single result regardless of their individual pattern structures. Using GEO, we optimize the batched query by setting a common reconstruction path for all the patterns in the batch. Figure 20 shows the results. As we can see, the rewritten batched queries generated by GEO result in a performance improvement of up to 64%.

*8.1.2 Individual Reconstructability.* In this case, the results for each pattern in the batched query are aggregated based on the individual pattern structures, and hence GEO optimizes the batched query by setting a unique reconstruction paths to each pattern in the batch. As shown in Figure 21, the rewritten batched queries generated by GEO result in a performance improvement of up to 62%. Compared to the collective reconstructabiltiy results from Figure 20, the execution times are higher. This is because the rewritten pattern batch, even though optimized, contains more patterns (e.g., for $APM_5$) and expensive patterns (e.g., for $APM_4$) compared to the batch generated for collective reconstructability in order to allow computing the subgraph results based on individual patterns.

## 8.2 Quasi-Clique Mining

Quasi-clique mining explores dense subgraphs. Quasi-cliques mining is useful across various applications, such as finding protein complexes or biologically relevant functional groups [Bu et al. 2003; Hu et al. 2005], social communities [Hopcroft et al. 2004; Li et al. 2014], and botnet crime analytics [Tanner et al. 2010; Weiss and Warner 2015]. We perform quasi-clique mining as defined in [Liu and Wong 2008]. A $\gamma$-quasi-clique of size k is a pattern $p$ with $k$ vertices, where each vertex is adjacent to at least $\gamma \times (k-1)$ vertices in $p$. Figure 22 shows size-4 and size-5 quasi-clique mining problems with two different $\gamma$ values. As expected, the translated pattern matching query batch for the higher $\gamma = 0.8$ contains fewer but dense patterns, while the batch for the lower $\gamma = 0.5$ is larger.

Focusing on exploring dense subgraphs, quasi-clique mining expects collective reconstructability. Figure 23 shows the results. GEO's optimized batch is up to 71% faster compared to the input query.

## 9 Related Work

*Pattern Matching Systems.* An array of parallel and distributed systems have been developed to efficiently execute pattern matching queries [Che et al. 2024; Chen and Qian 2022, 2023; Chen and Arvind 2022; Chen et al. 2021; Gui et al. 2023; Jamshidi et al. 2020; Jamshidi and Vora 2021; Mawhirter and Wu 2019]. Core to these systems are subgraph matching engines and sophisticated cost models to select an efficient execution plan. Geo provides such systems with equivalent yet more efficient pattern matching queries using the system's built-in cost model.

*Subgraph Matching Algorithms.* Considerable work has been done on subgraph matching [Ammar et al. 2018; Bhattarai et al. 2019; Bi et al. 2016; Han et al. 2019, 2013; Kim et al. 2016; Lai et al. 2015, 2016; Mhedhbi and Salihoglu 2019,?; Qiao et al. 2017; Ren et al. 2019; Shao et al. 2014; Shi et al. 2020; Wang et al. 2023; Yang et al. 2021a]. We refer readers to to a recent study for detailed comparison and evaluation of recent methods [Sun and Luo 2020]. Geo can optimize queries to be executed on any subgraph matching algorithm whose cost can be accurately modeled.

*Cost Estimation.* Subgraph matching algorithms and graph mining systems rely on cost estimation techniques to select an efficient execution plan from a combinatorial search space of possible plans. Individual works often develop tailor-made cost models, but there is also dedicated work on cardinality estimation, the crux of these cost models, which estimates the number of matches generated by a given step in an execution plan. There are dedicated subgraph matching cardinality estimation works [Chen et al. 2022; Hu and Motik 2024], as well as a rich literature of relational join cardinality estimation techniques [Cai et al. 2019; Kim et al. 2022; Wu et al. 2023] which can be applied to subgraph matching. However, a recent study shows these traditional methods are less effective than graph-specific techniques [Park et al. 2020]. Advancements in cost estimation allow Geo to more effectively optimize pattern matching queries, strengthening its contributions.

*Pattern Equivalences.* Prior works have explored optimizing subgraph counting through pattern equivalences exploiting structural relationships between patterns [Jamshidi et al. 2023; Marcus and Shavitt 2012; Melckenbeeck et al. 2017, 2019; Pinar et al. 2017; Zhang et al. 2020]. They do not provide a general framework that automatically applies user-defined equivalences for different graph mining problems. Subgraph Morphing [Jamshidi et al. 2023] automatically applies one class of pattern equivalences after a heuristic search through different candidate pattern sets, resulting in a potentially suboptimal pattern set. By contrast, Geo allows users to define custom pattern equivalence rules, optimizes across different rules, and explores the entire search space to guarantee a cost-optimal pattern set under certain conditions, utilizing these prior works.

*Canonicalization for Equivalence in Automated Programming Tools.* Cobbler [Lubin et al. 2024] used canonicalization functions in the interest of identifying Component-Based Refactorings. Cobbler takes a program as input, and will enumerate candidate programs derived from given components, until one is identified that is equivalent to the input. In this domain, canonicalization is done to enable quick equivalence checks and variable unification. Tensat [Yang et al. 2021b] performs canonicalizations as part of its e-graph construction as well, in the interests of shrinking the search space when performing code optimizations. Tensat specifically canonicalizes according to variable renames, and have a specific canonicalization-decanonicalization algorithm to ensure programs are correctly transformed. By contrast, we specifically do not ever decanonicalize, and have identified general requirements in which our canonicalization-centric approach works well.

*Query Optimization Systems Through Rewrites.* Query optimization has a long history, and there has been a recent flourishing of papers performing optimizations via program rewrites. Many of these optimization engines apply in relational data management systems. For example, Amazon Redshift integrates a rewrite-based optimizer into their query pipeline [Armenatzoglou et al. 2022].

A recent rewrite-based optimizer is `LearnedRewrite` [Zhou et al. 2021], which finds query optimizations via repeated applications of rewrite rules, where the order of rewrite rule application is determined via Monte Carlo Tree Search. Where our tool traverses equivalences via via `egg`'s equality saturation, one could imagine using `LearnedRewrite` MCTS instead. Our insights regarding canonicalizations and EmRec in pattern mining should generalize to other such systems. Rather than focus on building rewrite engines, other approaches like WeTune [Wang et al. 2022] aim to identify new rewrites. There is interesting future work in integrating rewrite discovery like that in WeTune to pattern mining equivalence discovery. Similar types of rewrite discovery engines have also been applied to other domains, like image processing [Newcomb et al. 2020]. An alternative approach is explored in SlabCity [Dong et al. 2023] is to do full query rewrites. Unlike rule-based query systems, SlabCity is not dependent on maintaining a list of rewrite rules, so it obviates the need for domain experts to spend time identifying and curating a high quality list of rewrite rules.

*E-Graphs for Optimization.* There has been a recent flourishing in using e-graphs for optimization, inspired in large part to the increased efficiently of the egg tool [Willsey et al. 2021]. E-Graphs have been used for optimizing circuits [Cheng et al. 2024; Coward et al. 2022] and tensor graphs [Yang et al. 2021b], for building faster term-rewriting compiler optimizers [Kourta et al. 2022], optimizing floating-point expression accuracy [Panchekha et al. 2015], and more. The RisingLight project uses e-graphs to build a SQL optimizer [Wang 2023]. This problem is similar to ours in that the costs is dependent on the underlying data, though the equivalences of the relational algebra are less complex than those of arbitrary graphs. In addition to these works, The Awesome E-Graphs page [Zucker and Goens 2024] lists and categorizes many tools built on and extensions to E-Graphs, and contains a more extensive list of various usages of e-graphs for optimization and more.

## 10 Future Work

There is interesting future work in (1) finding new types of rewrite rules for graph pattern mining optimizations, (2) applying EmRec in new domains to enable more optimizations, and (3) using canonicalized e-graphs to enable optimizations in other domains with complex equational theories.

*New Rewrite Rules for Graph Pattern Mining.* With this work, we have lowered the barrier for safely applying novel rewrites for pattern query optimization. The bar for applying such research is lower with Geo, which has enabled such equivalences to have broader application. Where previously such rewrites only benefitted the patterns they themselves identified, Geo lets these equivalences interact, letting the research apply in situations where such patterns can be used as sub-queries.

*Using EmRec for Batching Problems.* We believe EmRec can be applied to other domains in which problems with shared sub-problems within the same batch can reuse results. Previously, such broad applications were identified through common subexpression elimination. However, with EmRec, such optimizations can be identified completely through equality saturation. We think this can help identify optimizations like view and index reuse in SQL, or improving cache utilization.

*Applications of Canonicalized E-Graphs.* We think canonicalized e-graphs can be useful in situations where a language has embedded structures with complex equational theories, yet simple canonicalization functions. Regular expressions permit simple canonicalizations, but their equational theory is infinite [Redko 1964]. Languages involving regular expressions, like parsing languages [Levine et al. 1992] or lens languages [Foster 2009], are amenable to optimizations via canonicalized e-graphs.

## 11 Conclusion

In this paper, we introduced Geo, a tool for building optimized pattern mining queries. Geo enables domain experts to build rewrites without worrying about minute details. Geo uses a combination of e-graphs and canonicalization to enable efficient exploration of the query equivalences, and under certain conditions it is guaranteed to return a minimal-cost query. In addition to increasing the expressivity of our language, developing flexible notions of reconstructability enables further optimization. We thoroughly evaluated Geo and ran it on two graph mining case studies, and find that it identifies highly optimized queries for these problems.

## Acknowledgements

We thank anonymous reviewers for their valuable and thorough feedback. This work is supported by the Natural Sciences and Engineering Research Council of Canada.

## Data-Availability

An artifact exists for this work [Yousefian et al. 2026]. This artifact is distributed as a docker image, and includes a README with step-by-step instructions.

## A E-Graph Formalism

An *e-graph* is a data structure that compactly represents sets of expressions, grouping equivalent terms into *e-classes*. Each e-class is a set of *e-nodes*, where each e-node represents a specific expression or part of an expression according to the language's grammar. An e-node consists of an operator and references to other e-classes, thus creating a graph-based structure for representing terms.

Formally, an e-graph $E = (U, M, H)$ includes:

(1) A *union-find structure* $U$, which maintains equivalence relations among e-class identifiers.
(2) An *e-class map* $M$, which associates e-class identifiers with sets of e-nodes that belong to each class.
(3) A *hash-consing map* $H$, which maps canonical e-nodes to e-class identifiers, ensuring unique representations within the e-graph.

Canonical e-nodes are e-nodes where the e-class ids are the canonical representatives in the union-find data structure. It is fast and easy to transform an e-node into a canonical e-node. Thus, for simplicity (and to avoid a namespace collision), we will be treating the hash-consing map $H$ as simply a mapping from e-nodes to e-class identifiers, rather than a transformer from canonical e-nodes to e-class identifiers.

The Representation of a term $t$ in an e-graph is defined recursively as follows:

(1) An e-graph $E$ represents a term $t$ if there exists at least one e-class $ec \in M$ such that $ec$ represents $t$.
$$Rep(E, t) \iff \exists ec \in M \text{ such that } Rep(ec, t)$$
(2) An e-class $ec$ represents a term $t$ if there exists at least one e-node $n \in ec$ such that $n$ represents $t$.
$$Rep(ec, t) \iff \exists n \in c \text{ such that } Rep(n, t)$$
(3) An e-node of the form $f(a_1, \ldots, a_n)$ represents a term $t = f(t_1, \ldots, t_n)$ if $f$ is the function symbol of $t$, and each e-class $M[a_i]$ represents the corresponding subterm $t_i$.
$$Rep(c(a_1, \ldots, a_n), c(t_1, \ldots, t_n)) \iff \forall i \in \{1, \ldots, n\}, Rep(M[a_i], t_i)$$

One can ensure a term is represented by an e-graph by adding that term to the e-graph (represented as $E \cup \{t\}$. This is done recursively. First, if $t = c(t_1, \ldots, t_n)$ let $(U', M', H') = E' = E \bigcup_i \{t_i\}$. Then, let $t_i \in M[a_i]$ for all $i$. If $c(a_i, \ldots, a_n)$ is not in the keys of $H'$, create a new e-class identifier $a$, let $M'' = M' \cup a \mapsto \{c(a_i, \ldots, a_n)\}$, let $H'' = H \cup c(a_i, \ldots, a_n) \mapsto a$. $E \cup \{t\} = (U', M'', H'')$. If $c(a_i, \ldots, a_n)$ is in the keys of $H'$, $E \cup \{t\} = E'$.

For a rewrite rule $(m, f)$, the process of rule application in an e-graph is as follows:

(1) *Match Identification:* Let $\sigma$ denote a substitution for variables in $m$ such that $m[\sigma]$ matches an e-class $c$ within $E$. We denote this by
$$ec \vdash m[\sigma] \iff \exists \sigma \text{ such that } Rep(ec, m[\sigma])$$
(2) *Rule Application:* When a match $ec \vdash m[\sigma]$ is found, the term $f(\sigma)$ is added to the e-graph. The e-class containing $f(\sigma)$ is then merged with $ec$, unifying their representations in $E$.
$$E := E \cup \{f(\sigma)\} \text{ and merge}(ec, \text{class of } f(\sigma))$$
(3) *Congruence Recovery:* For all e-nodes $n = c(a_1, \ldots, a_i, \ldots, a_n)$ and $n' = c(a_1, \ldots, a_i', \ldots, a_n)$ where $H[a_i] = m[\sigma]$ and $H[a_i'] = f(\sigma))$, merge$(M[H[n]], M[H[n']])$. Recursively perform invariant recovery.

*Definition A.1.* Given a set of rewrites $R$ and an e-graph $E = (U, M, H)$, we define $R(E)$ as the e-graph $E' = (U', M', H')$ where:

(1) Identify all matches $ec \vdash m[\sigma]$ for any $(m, f) \in R$
(2) Apply *Rule Application* for every $f(\sigma)$ in the matches.
(3) Perform *Congruence Recovery* for every $m[\sigma] \mapsto f(\sigma)$ applied.

Notably, this means that running $R(E)$ corresponds exactly to one egg equality saturation. $R^*(E)$ is the result of running $R$ on $E$ until it reaches a fixpoint (in other words, saturation).

We would like a way to compare the two union-find data structures of e-graphs.

*Definition A.2.* Given an E-Graph $E = (U, M, H)$, the induced equivalence $\equiv_E$ is defined as:

(1) If $t$ is not represented by $E$, then $t \neg \equiv_E t'$ for any $t' \neq t$.
(2) If $t$ and $t'$ are both represented by $E$, $t \equiv_E t'$ there exists some $ec$ such that $Rep(ec, t)$ and $Rep(ec, t')$.

*Definition A.3.* A union-find data structure $U$ is *finer* than a union-find data structure $U'$ if $U.find(x) = U.find(y)$ implies that $U'.find(x) = U'.find(y)$. If $U$ is finer than $U'$, then $U'$ is coarser than $U$.

Being a coarser union-find data structure is preferable. This is because one can perform a "find" operation on an e-graph. Given an e-graph $E$ and a cost metric $C$, $E.find(t)$ will get a term $t'$ such that $t \equiv_E t'$ and for all $t''$, if $t \equiv_E t''$ then $C(t') \leq C(t'')$.

## A.1 Canonicalized E-Graphs

Canonicalized E-Graphs are essentially a reinterpretation of e-graphs, where the induced equivalence is done on the presumption of looking up exclusively canonicalized terms. Typically, only canonical terms are represented by the e-graph, enabling a coarser induced equivalence relation, despite requiring fewer rewrites.

These canonicalized e-graphs have a slightly different notion of the equivalence classes. Namely, two terms are equivalent if their canonical representatives are equivalent in the standard equivalence classes. This aims to capture the notion that, under a canonicalized e-graph, one performs the e-graph *find* and union operations on the canonicalized form of a term, not the original term.

*Definition A.4.* Given an E-Graph $E = (U, M, H)$ and a canonicalization $\varphi$, we define equivalence $t_1 \equiv_E^{\varphi} t_2$ as $\varphi(t_1) \equiv_E \varphi(t_2)$

A primary benefit of canonicalized e-graphs are that they enable the elision of rules dealing with the equivalences of canonicalization. Essentially, under the assumption that the rewrite rules respect the canonicalization, one can completely omit a subset of the rewrite rules, and the induced equivalence remains coarser. This ability to omit rewrite rules is key in establishing correctness of our overall algorithm.

# B Proofs

Theorem B.1. *Restatement of Theorem 4.12. Let $Q$ be a graph query. For all graphs $G$, $[\![Q]\!]_G = [\![\varphi_{PMQ}(Q)]\!]_G$.*

Proof. We strengthen the IH: For all graphs $G$, $[\![Q]\!]_G = [\![\varphi_{PMQ}(Q)]\!]_G$. For all Patterns $P$, $[\![P]\!] = [\![\varphi_{PMQ}(P)]\!]$. For all Provenance Scalers $\Pi$, $[\![\Pi]\!] = [\![\varphi_{PMQ}(\Pi)]\!]$.

We prove this by induction over the structure of the query.

**[Case 1:]** $Q = \texttt{Union}(Q_1, Q_2)$. $[\![\varphi_{PMQ}(\texttt{Union}(Q_1, Q_2))]\!]_G = [\![(\texttt{Union}(\varphi_{PMQ}(Q_1), \varphi_{PMQ}(Q_2))]\!]_G$ by the definition of $\varphi_{PMQ}$. Let $\pi$ be an arbitrary provenance variable. By definition, $[\![(\texttt{Union}(\varphi_{PMQ}(Q_1), \varphi_{PMQ}(Q_2))]\!]_G(\pi) = [\![\varphi_{PMQ}(Q_1)]\!]_G(\pi) + [\![\varphi_{PMQ}(Q_2)]\!]_G(\pi)$, which by IH equals $[\![Q_1]\!]_G(\pi) + [\![Q_2]\!]_G(\pi) = [\![\texttt{Union}(Q_1, Q_2)]\!](\pi)$, as desired.

**[Case 2:]** $Q = \texttt{Count}(\Pi, Q')$. $[\![\varphi_{PMQ}(\texttt{Count}(\Pi, Q'))]\!]_G = [\![(\texttt{Count}(\varphi_{PMQ}(\Pi), \varphi_{PMQ}(Q'))]\!]_G$ by the definition of $\varphi_{PMQ}$. Let $\pi$ be an arbitrary provenance variable. By definition,

$$\begin{array}{llcl}
\text{Queries} & Q & ::= & \texttt{Union}(Q_1,Q_2) \\
& & | & \texttt{Union}(\mathcal{X},Q_2) \\
& & | & \texttt{Union}(Q_1,\mathcal{X}) \\
& & | & \texttt{Union}(\mathcal{X},\mathcal{X}) \\
& & | & \texttt{Count}(\Pi,Q) \\
& & | & \texttt{Count}(\mathcal{X},Q) \\
& & | & \texttt{Count}(\Pi,\mathcal{X}) \\
& & | & \texttt{Count}(\mathcal{X},\mathcal{X}) \\
& & | & \texttt{Pattern}(P,F) \\
& & | & \texttt{Pattern}(\mathcal{X},F) \\
& & | & \texttt{Pattern}(\mathcal{X},\mathcal{X}) \\
\text{Patterns} & P & ::= & \texttt{Edge}(v_1,v_2)\ P \\
& & | & \texttt{Anti-Edge}(v_1,v_2)\ P \\
& & | & \cdot \\
\text{Reconstruction Path} & \Pi & ::= & \Pi + \Pi \\
& & | & \mathcal{X} + \Pi \\
& & | & \Pi + \mathcal{X} \\
& & | & \mathcal{X} + \mathcal{X} \\
& & | & (\pi, x) \\
& & | & (\mathcal{X}, x) \\
& & | & (\mathcal{X}, \mathcal{X}) \\
& & | & (\pi, \mathcal{X})
\end{array}$$

Fig. 24. Formal grammar of opaque matchers.

$[\![(\texttt{Count}(\varphi_{PMQ}(\Pi),\varphi_{PMQ}(Q'))]\!]_G(\pi) = [\![\varphi_{PMQ}(\Pi)]\!](\pi_1) \times [\![\varphi_{PMQ}(Q')]\!]_G(\pi_2)$ where $\pi_1\pi_2 = \pi$, which by IH equals $[\![\Pi]\!](\pi_1) \times [\![Q']\!]_G(\pi_2) = [\![\texttt{Count}(\Pi,Q')]\!](\pi)$, as desired.
**[Case 3:]** $Q = \texttt{Pattern}(P,F)$. $[\![\varphi_{PMQ}(\texttt{Pattern}(P,F))]\!]_G = [\![(\texttt{Pattern}(\varphi_{PMQ}(P),F)]\!]_G$ by the definition of $\varphi_{PMQ}$ (the $F$ variables are considered semantic and ignored by the canonizer). Let $\pi$ be an arbitrary provenance variable. If $\pi$ is not 1, it maps to 0, as desired, so let $\pi$ be 1. By definition, $[\![(\texttt{Pattern}(\varphi_{PMQ}(P),F)]\!]_G(1) = \|\{\sigma \mid [\![\varphi_{PMQ}(P)]\!](G,\sigma) \wedge F(\sigma)\}\|$, which by IH equals $\|\{\sigma \mid [\![P]\!](G,\sigma) \wedge F(\sigma)\}\| = [\![(\texttt{Pattern}(P,F)]\!]_G(1)$, as desired.
**[Case 4:]** $P$ is a pattern. Since $P$ is a pattern, $\varphi_{PMQ}(P) = \varphi_{Graph}(P)$. By assumption, $[\![\varphi_{Graph}(P)]\!] = [\![P]\!]$.
**[Case 5:]** $\Pi = \Pi_1 + \Pi_2$. $[\![\varphi_{PMQ}(\Pi_1 + \Pi_2)]\!] = [\![\varphi_{PMQ}(\Pi_1) + \varphi_{PMQ}(\Pi_2)]\!]$ by the definition of $\varphi_{PMQ}$. Let $\pi$ be an arbitrary provenance variable.

By definition, $[\![\varphi_{PMQ}(\Pi_1) + \varphi_{PMQ}(\Pi_2)]\!](\pi) = [\![\varphi_{PMQ}(\Pi_1)]\!](\pi) + [\![\varphi_{PMQ}(\Pi_2)]\!](\pi)$, which by IH equals $[\![\Pi_1]\!](\pi) + [\![\Pi_2]\!](\pi) = [\![\Pi_1 + \Pi_2]\!](\pi)$, as desired.
**[Case 6:]** $\Pi = (\pi, x))$. $\varphi_{PMQ}((\pi,x)) = (\pi,x)$, so we are done.

□

*Definition B.2.* More formal definition of Definition 4.13. A pattern is opaque if it conforms to the grammar described in Figure 24

Theorem B.3. *Restatement of Theorem 4.14. Let $(m, f)$ be a rewrite rule. If $m$ treats patterns opaquely, then the rewrite rule $(\varphi_{PMQ}(m), \varphi_{PMV} \circ f \circ \varphi_{PMQ})$ respects $\varphi$.*

Proof. This proof proceeds via two lemmas.

First, we prove that if $m$ is opaque, then $m[\sigma] = t$ implies that $\varphi(m)[\varphi(\sigma)] = \varphi(t)$. Once we have proven this, as $\varphi(f(\varphi(\varphi(\sigma)))) = \varphi(f(\varphi(\sigma)))$, and the output is always canonicalized, we know that $(\varphi_{PMQ}(m), \varphi_{PMV} \circ f \circ \varphi_{PMQ})$ respects $\varphi$, as desired.

So we simply must prove that that if $m$ is opaque, then $m[\sigma] = t$ implies that $\varphi(m)[\varphi(\sigma)] = \varphi(t)$. We prove this by induction over the grammar of Figure 24.

We will do only a few illustrative cases, and the interesting cases.

If $m = \texttt{Union}(Q_1, Q_2)$ and $m[\sigma] = t$, then $t = \texttt{Union}(t_1, t_2)$ where $Q_1[\sigma] = t_1$ and $Q_2[\sigma] = t_2$. By IH, $\varphi(Q_1)[\varphi(\sigma)] = \varphi(t_1)$ and $\varphi(Q_2)[\varphi(\sigma)] = \varphi(t_2)$. As $\varphi(\texttt{Union}(t_1, t_2)) = (\texttt{Union}(\varphi(t_1), \varphi(t_2)))$ (and similarly with $m$), $\varphi(\texttt{Union}(Q_1, Q_2))$ matches $\varphi(\texttt{Union}(t_1, t_2))$.

If $m = \texttt{Union}(X_1, X_2)$ and $m[\sigma] = t$, then $t = \texttt{Union}(t_1, t_2)$. This means that $\sigma(X_1) = t_1$ and $\sigma(X_2) = t_2$. This means $\varphi(\sigma(X_1)) = \varphi(t_1)$ and $\varphi(\sigma(X_2)) = \varphi(t_2)$. As $\varphi(\texttt{Union}(t_1, t_2)) = (\texttt{Union}(\varphi(t_1), \varphi(t_2)))$, and similarly for $m$, we know that $\varphi(\texttt{Union}(\mathcal{X}_\infty, \mathcal{X}_\in))[\varphi(\sigma)] = \varphi(t)$ as desired.

If $m = \texttt{Pattern}(P, F)$ and $m[\sigma] = t$, then $t = \texttt{Pattern}(P, F)$, as there are no meta-variables, and so $m = t$, so $\varphi(\texttt{Pattern}(P, F))[\sigma] = \varphi(\texttt{Pattern}(P, t_2))$.

If $m = \texttt{Pattern}(X, F)$ and $m[\sigma] = t$, then $t = \texttt{Pattern}(P, t_2)$. As $\varphi(\texttt{Pattern}(X, F)) = \texttt{Pattern}(\mathcal{X}, F)$, we know $\sigma(X) = P$. This means $\varphi(\sigma(X)) = \varphi(P)$. This means that as $\varphi(\texttt{Pattern}(X, F))[\varphi\sigma] = \varphi(t)$ matches any $t$ we know $\varphi(\texttt{Pattern}(X, F))[\varphi(\sigma)] = \varphi(\texttt{Pattern}(P, F))$.

The remaining cases are uninteresting, and follow the same structure as the first two cases. ◻

Lemma B.4. *If $t_1 \to_R t_2$, and $t_1$ is represented by $E = (U, M, H)$, then $t_1 \equiv_{R(E)} t_2$*

Proof. We proceed by induction over the structure of $t$.

There are two cases:

**[Case 1:]** $(m, f) \in R$ and there exists some $\sigma$ such that $m[\sigma] = t_1$ and $f(\sigma) = t_2$

This is immediate, as the e-node identifier $a_1$ such that $M[a_1]$ represents $t_1$ is merged with the e-node identifier $a_2$ such that $M[a_2]$ represents $t_2$.

**[Case 2:]** $t_1 = c(u_1, \ldots, u_n)$ and $t_2 = c(u'_1, \ldots, u'_n)$ and $u_i \to_R u'_i$ or $u_i = u'_i$

By IH, $u_i \equiv_E u'_i$ for all $i$. By *Congruence Recovery*, $t_1 \equiv_E t_{1i} = c(u_1, \ldots, u'_i, \ldots, u_n)$. By transitivity, $t_1 \equiv_E t_2$. ◻

Lemma B.5. *Let $R$ respect $\varphi$. Let $\equiv_\varphi = \equiv_T$. If $t_1 \to_{R \cup T} t_2$, then $\varphi(t_1) \to_R \varphi(t_2)$.*

Proof. By induction over the derivation of $t \to_{R \cup T} t'$.

**Case 1:** $(m, f) \in R$ and there exists some $\sigma$ such that $m[\sigma] = t_1$ and $f(\sigma) = t_2$.

As $R$ respects $\varphi$, we know that $\varphi(t) \to_R \varphi(t')$

**Case 2:** $(m, f) \in T$ and there exists some $\sigma$ such that $m[\sigma] = t_1$ and $f(\sigma) = t_2$.

As $\equiv_\varphi = \equiv_T$, $\varphi(t_1) = \varphi(t_2)$, so $\varphi(t_1) \to_R \varphi(t_2)$

**Case 3:** If $t_1 = c(u_1, \ldots, u_n)$ and $t_2 = c(u'_1, \ldots, u'_n)$ and $u_i \to_{R \cup T} u'_i$ or $u_i = u'_i$.

By IH, $\varphi(u_i) \to_R \varphi(u'_i)$. Thus, $c(\varphi(u_1), \ldots, \varphi(u_n)) \to_R c(\varphi(u'_1), \ldots, \varphi(u'_n))$, so we know $\varphi(c(\varphi(u_1), \ldots, \varphi(u_n))) \to_R \varphi(c(\varphi(u'_1), \ldots, \varphi(u'_n)))$. As $\forall x. \varphi(x) = x$, we know $\varphi(c(\varphi(u_1), \ldots, \varphi(u_n))) = \varphi(c(u_1, \ldots, u_n))$ and $\varphi(c(\varphi(u'_1), \ldots, \varphi(u'_n))) = \varphi(c(u'_1, \ldots, u'_n))$, so $\varphi(c(u_1, \ldots, u_n)) \to_R \varphi(c(u'_1, \ldots, u'_n))$ ◻

Lemma B.6. *Let $R$ respect $\varphi$, and $E$ be a e-graph such that $R(E) = E$. Let $\equiv_\varphi = \to^*_T$. If $t \to^*_{R \cup T} t'$, then $\varphi(t) \equiv_E \varphi(t')$.*

Proof. By induction over the steps of $\to^*_{R \cup T} t'$.

**Case 1:** $t = t'$

Trivial

**Case 2:** $t \to^*_{R \cup T} t''$ and $t'' \to_{R \cup T} t'$.

By IH, $\varphi(t'') \equiv_E \varphi(t)$. As $\varphi(t)$ is represented by $E$, this means that $\varphi(t'')$ is represented by $E$. As $t'' \to_{R \cup T} t'$, by Lemma B.5, $\varphi(t'') \to_R \varphi(t')$. By Lemma B.4, $\varphi(t'') \equiv_{R(E)} \varphi(t')$, so $\varphi(t'') \equiv_E \varphi(t')$. ◻

Lemma B.7. *Let $R$ respect $\varphi$, and $E_0$ be a e-graph such that $equiv_E = (=)$. Let $E = R^n(E_0)$. Let $\equiv_\varphi = \to^*_T$. If $\varphi(t) \equiv_E \varphi(t')$, then $t \to^*_{R \cup T} t'$.*

Proof. By induction on n.

**Case 1:** $n = 0$

$t \rightarrow^*_{R\cup T} t$

**Case 2:** $n = n' + 1$

Let $\varphi(t) \equiv_E \varphi(t')$. This means $\varphi(t)$ and $\varphi(t')$ are represented by the same e-class $ec$. We do not prove the soundness of e-graphs in this paper (existing result from Nelson and Oppen [Nelson and Oppen 1980]), so we take it as a given that $\varphi(t) \rightarrow^*_R \varphi(t')$. Furthermore, $t \rightarrow^*_T \varphi(t)$ as $\equiv_\varphi = \rightarrow^*_T$. Furthermore, $\varphi(t') \rightarrow^*_T t'$ as $\equiv_\varphi = \rightarrow^*_T$. Thus, $t \rightarrow^*_{R\cup T} t'$.

Let $\varphi(t) \equiv_E \varphi(t')$. This means there exists some e-class $ec$ such that $ec$ represents □

Theorem B.8. *Restatement of Theorem 5.1. Let $R$ respect $\varphi_{GPM}$ and $S = \{R'(E) = E\}$ and Algorithm 1 returns a query $Q'$. That $Q'$ is a minimal cost query that satisfies $Q \rightarrow^*_{R\cup A} Q'$.*

Proof. By Lemma B.7, we know that $Q'$ satisfies $Q \rightarrow^*_{R\cup A} Q'$.

Furthermore, let $Q \rightarrow_{R\cup A} Q''$. We know that $C(\varphi(Q'')) \leq$ Then, by Lemma B.6, $Q \equiv^{\varphi_{GPM}}_E Q''$. So for all other $Q''$, we know that $\varphi(Q'') \equiv^{\varphi}_E (Q')$, so $C(Q') \leq C(\varphi(Q''))$, by the definition of *find*, and $C(\varphi(Q'')) \leq C(Q'')$ as $C$ respects $R \cup A$. Thus, by transitivity, $Q'$ has a lower cost than $Q''$. □